\documentclass[aps,prb,twocolumn,letterpaper]{revtex4}
\usepackage{graphicx}
\usepackage{dcolumn}
\usepackage{amsmath}
\usepackage{amssymb}
\usepackage{subfigure}
\usepackage{float}
\usepackage{epstopdf}
\usepackage{natbib}
\usepackage{color}

\DeclareMathOperator{\arccosh}{arccosh}
\DeclareMathOperator{\csch}{csch}
\DeclareMathOperator{\sech}{sech}
\begin{document}

\newcommand{\dtau}{\partial_\tau}

\newcommand{\ket}[1]{\left| #1 \right\rangle}
\newcommand{\bra}[1]{\left\langle #1 \right|}

\title{Revisiting $2\pi$ phase slip suppression in topological Josephson junctions}
\author{Rosa Rodr\'{i}guez-Mota$^1$, Smitha Vishveshwara$^2$ and T. Pereg-Barnea$^{1,3}$}
\affiliation{$^1$Department of Physics and the Centre for Physics of Materials, McGill University, Montreal, Quebec, Canada H3A 2T8\\
$^2$Department of Physics, University of Illinois at Urbana-Champaign, Urbana, Illinois 61801-3080, USA\\
$^3$Department of Condensed Matter Physics, Weizmann Institute of Science, Rehovot, 76100, Israel}
\date{\today}

\newcommand{\etalcomma}{\textit{et al.,\ }}

\definecolor{pink}{RGB}{228,0,124}
\definecolor{green}{RGB}{0,153,76}
\definecolor{moradito}{RGB}{153,31,255}
\definecolor{amarillo}{RGB}{255,153,51}

\begin{abstract}

Current state of the art devices for detecting and manipulating Majorana fermions commonly consist of networks of Majorana wires and tunnel junctions. We study a key ingredient of these networks -- a topological Josephson junction with charging energy -- and pinpoint crucial features for device implementation. The phase dependent tunneling term contains both the usual $2\pi$-periodic Josephson term and a $4\pi$-periodic Majorana tunneling term representing the coupling between Majoranas on both sides of the junction. In non-topological junctions when the charging energy is small compared to the Josephson tunneling scale the low energy physics is described by $2\pi$ phase slips. By contrast, in a topological junction, due to the $4\pi$ periodicity of the tunneling term it is usually expected that only $4\pi$ phase slips are possible while $2\pi$ phase slips are suppressed. However, we find that if the ratio between the strengths of the Majorana assisted tunneling and the Josephson tunneling is small, as is likely to be the case for many setups, $2\pi$ phase slips occur and may even dominate the low energy physics. In this limit one can view the $4\pi$ phase slips as a pair of $2\pi$ phase slips with arbitrarily large separation. We provide an effective descriptions of the system in terms of $2\pi$ and $4\pi$ phase slips valid for all values of the tunneling ratio. Comparing the spectrum of the effective models with numerical simulations we determine the cross-over between the $4\pi$ phase slip regime to $2\pi$ phase slip dominated regime. We also discuss the role of the charging energy as well as the implications of our results on the dissipative phase transitions expected in such a system. 
\end{abstract}

\maketitle

\section{Introduction}

In recent years, extensive scientific efforts have been invested to understand, realize and manipulate topological states in condensed matter.\cite{doi:10.1146/annurev-conmatphys-031214-014740,RevModPhys.89.040502}
Particularly, topological superconducting wires~\cite{PhysRevLett.105.077001,PhysRevLett.105.177002}, which are constructed using systems with strong spin-orbit coupling, induced Cooper pairing and a Zeeman field, have gathered much attention~\cite{aguadoreview}. 
Interest in these Majorana wires is motivated by the possibility of using the non-Abelian nature of Majorana modes for quantum computation schemes~\cite{Sarma2015,1063-7869-44-10S-S29} and is sustained by encouraging experimental results~\cite{Albrecht2016,2016arXiv160304069Z,Deng1557,PhysRevLett.110.126406,Das2012,PhysRevB.87.241401,doi:10.1021/nl303758w,Mourik1003,Rokhinson2012}.   
Hence, networks of Majorana wires have been proposed as tools to manipulate Majorana modes for quantum information purposes~\cite{PhysRevB.88.035121,PhysRevLett.106.090503,PhysRevLett.106.130505,1367-2630-14-3-035019,PhysRevX.6.031016,PhysRevB.95.235305} or to create more exotic matter~\cite{1367-2630-14-12-125018,PhysRevB.96.121119}. In this work, we study a phenomenon commonly relevant to this type of networks, charge induced quantum fluctuations in topological Josephson junctions.

In a superconductor, the charge is conjugate to the order parameter phase and charging effects induce quantum phase fluctuations.\cite{Likharev1985,bloch} In a non-topological Josephson junction, tunneling processes are known as phase slips and are essentially $2\pi$ jumps in the phase difference between the superconductors. The delocalization of the phase induced by these fluctuations can be prevented by dissipation. As a result, Josephson junctions present a dissipative phase transition \cite{Panyukov1988,SCHON1990237}. In a topological junction which is made of two topological superconductors there are Majorana modes at both edges of the junction. The presence of these modes leads to coherent single particle tunneling between the superconductors, commonly referred to as the $4\pi$ periodic Josephson effect.~\citep{1063-7869-44-10S-S29,Kwon2004,PhysRevB.79.161408,:/content/aip/journal/ltp/30/7/10.1063/1.1789931,PhysRevLett.105.077001,PhysRevLett.105.177002,Badiane2013840} The change of periodicity in the overall tunneling current suppresses $2\pi$ phase slips in topological Josephson junctions.~\cite{PhysRevB.87.064506} Both the $2\pi$ phase slip suppression~\cite{PhysRevB.87.064506,2017arXiv170600576Z,PhysRevLett.112.247001}, and its effects on the dissipative phase transition~\cite{PhysRevLett.112.247001} have been proposed as a probe for topological superconductivity. Most studies of $2\pi$ phase slip suppression focus on having a sufficiently strong single particle tunneling.~\cite{PhysRevB.87.064506,2017arXiv170600576Z,PhysRevLett.112.247001} This is despite the fact that the single particle tunneling may be a small component of the overall tunneling current, as is the case for 3D topological insulator based Josephson junctions~\cite{Kurter2015,4piTISC}. As a result, there are currently no studies which describe the $2\pi$ phase slip suppression throughout the transition from a non-topological to a topological junction.

\begin{figure}
\centering
\subfigure[~]{
\includegraphics[width=0.8\linewidth]{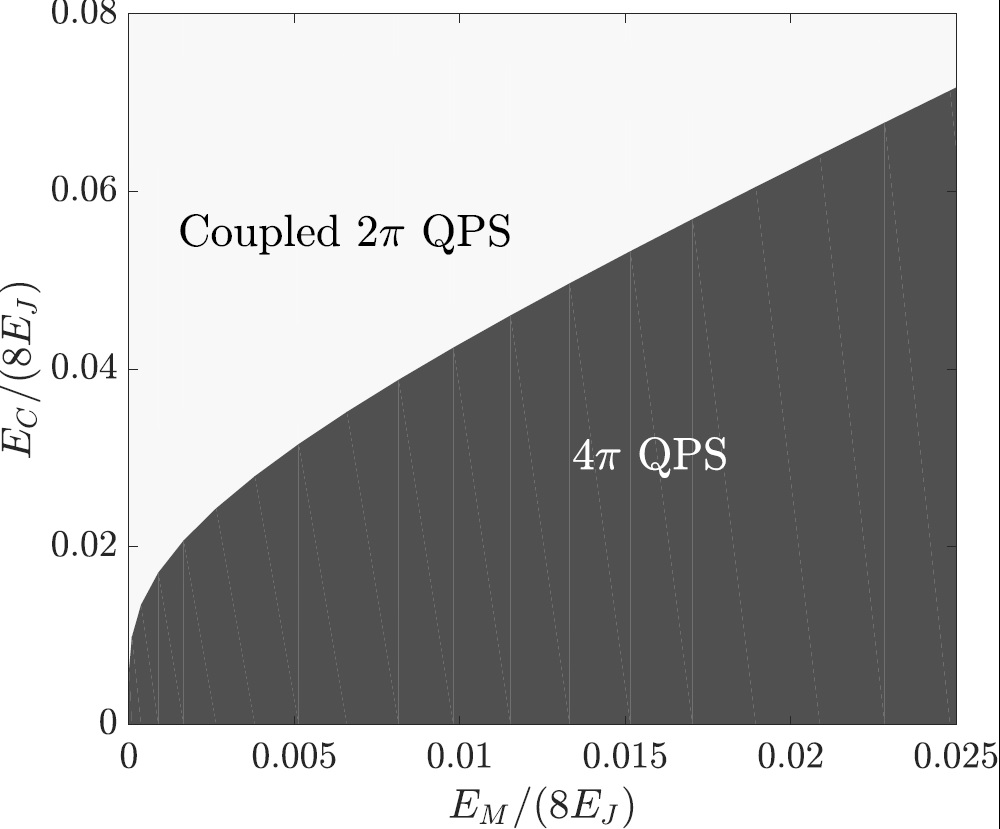}\label{fig:Main}}
\subfigure[$E_M/(8E_J)>1$]{
\includegraphics[width=0.4\linewidth]{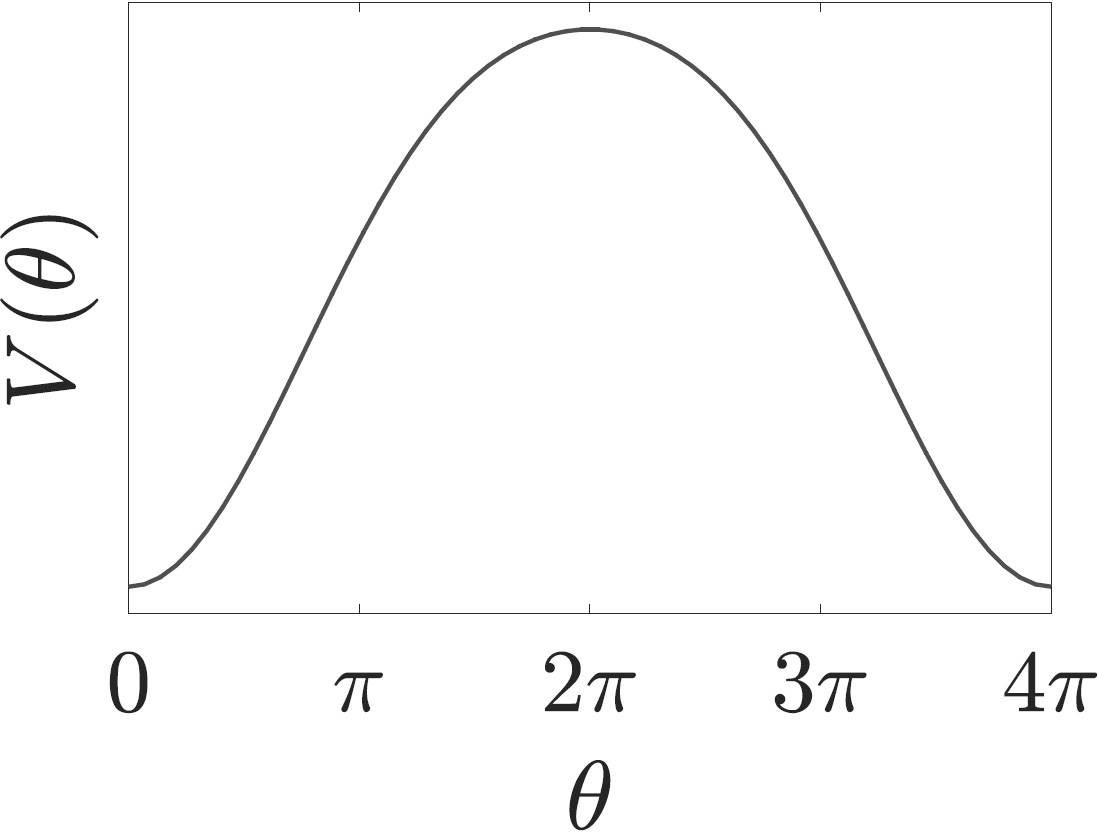}\label{fig:V1min}}
\subfigure[$E_M/(8E_J)<1$]{
\includegraphics[width=0.4\linewidth]{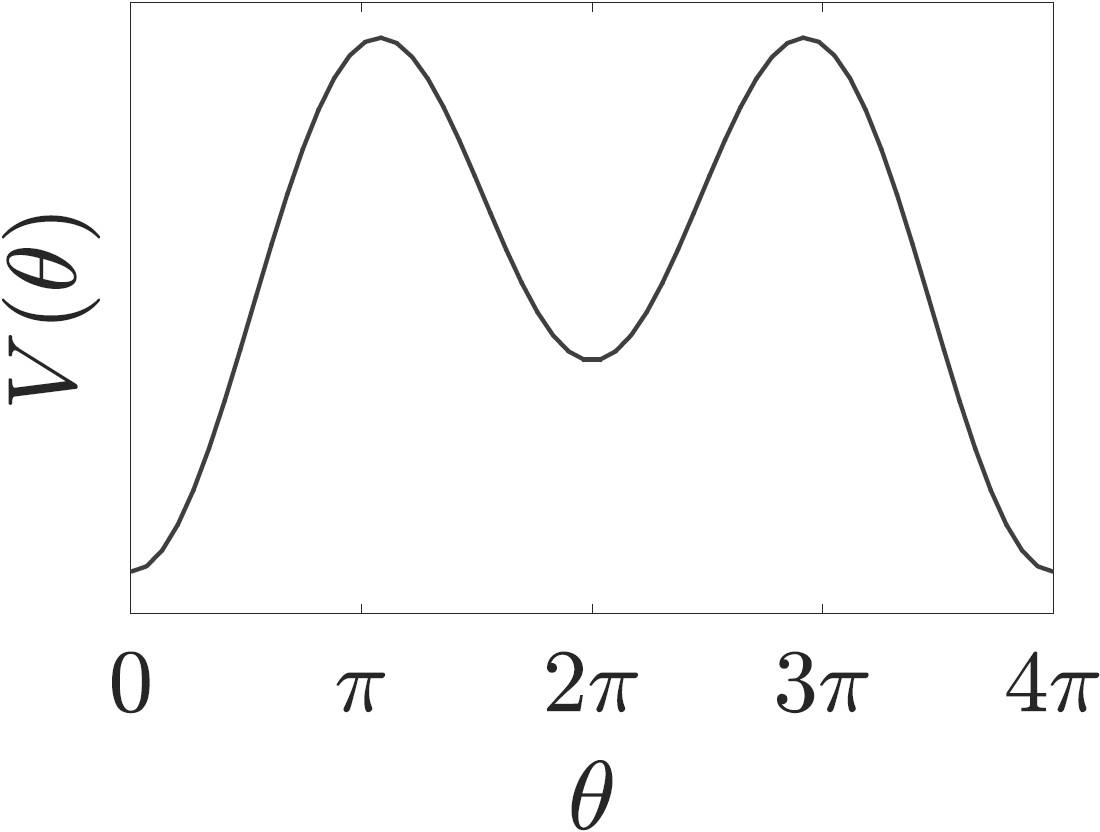}\label{fig:V2min}}
\caption{Depending on the relative strength between the single-particle (set by $E_M$) and the pair tunneling (set by $E_J$) the potential of the topological Josephson junction may be minimized when: (b) the phase difference across the junction is an integer multiple of $4\pi$ only, or (c) the phase difference across the junction is any integer multiple of $2\pi$. In (c) the minima at odd $2\pi$ are local minima. (a) In (c), the strength of phase fluctuations (set by the charging energy $E_C$) determines whether oscillations around the local minima contribute to the ground-state (Coupled $2\pi$ QPS) or not ($4\pi$ QPS).  The cross-over is found numerically by evaluating the relative accuracy of the $4\pi$ and $2\pi$ phase slip scenarios.}
\label{fig:MainResults}
\end{figure}

In this work, we develop a theory for the effect of charging induced quantum fluctuations in the low energy spectrum of a topological Josephson junction, valid for any ratio of the single particle and the Cooper pair tunneling. Our results show that a description of the low energy physics of the topological junction in terms of $4\pi$ phase slips alone may be insufficient if the strength of the $4\pi$ periodic tunneling is small. In the presence of both $2\pi$ and $4\pi$ periodic components of the tunneling current, the potential energy of the junction as a function of the phase difference between the superconductors, $\theta$, may have one or two minima in $[0,4\pi)$ (see Figs.~\ref{fig:V1min} and~\ref{fig:V2min}). If only one minimum exists, the description in terms of $4\pi$ phase slips is valid as long as the phase fluctuations are small. In the presence of two minima, this description may break down even for small phase fluctuations if they are relatively large compared to the strength of the $4\pi$ periodic tunneling. In this case, a description of the junction in terms of coupled $2\pi$ phase slips is more appropriate. This is shown schematically in Fig.~\ref{fig:Main} where $E_J$ and $E_M$ correspond to the energy scale of the $2\pi$ and $4\pi$ periodic tunneling, respectively, and $E_C$ to the strength of the phase fluctuations. The junction potential has only one minimum for $E_M>8E_J$ and two otherwise.

We treat the appearance of phase slips in the topological Josephson junction in two ways.
First, we calculate the phase slip probability using a semiclassical method.
In this case, a path integral between a state with $\theta=0$ in the distant past and $\theta=4\pi$ in the distant future describes the phase slip process. 
With Majorana assisted tunneling, the junction Lagrangian corresponds to the double-sine-Gordon model which presents a $4\pi$ instanton. 
We calculate the phase slip probability up to Gaussian fluctuations around this instanton.
In the small Majorana tunneling regime we assume dominance of $2\pi$ phase slips and calculate their probability using a method for asymmetric barriers~\cite{PhysRevA.86.012106}.
Secondly, we solve the problem numerically in a truncated Hilbert space.
The numerics give us the ground state energy of the junction as a function of a tunning parameter which we can compare with the spectrum expected for the $2\pi$ and $4\pi$ phase slip scenarios.
This gives us a regime of validity for either scenario and therefore a cross-over between the two behaviors, as depicted in Fig.~\ref{fig:Main}.

This paper is organized as follows. In Sec.~\ref{sec:review}, we give a small review of the effects of charging induced phase fluctuations in Josephson junctions. This is followed by a qualitative discussion of the effects of phase fluctuations for different regimes of a topological Josephson junction in Sec.~\ref{sec:Effects}. The main results are stated in Sec.~\ref{sec:Effective} where we introduce low energy effective models of topological Josephson junctions. In Sec.~\ref{sec:Dis} we discuss the implications of our results on the dissipative phase transition. Our conclusions are stated in Sec.~\ref{sec:conclusion}.
 
\section{Review of the effects of quantum phase slips in Josephson junctions}
\label{sec:review}
We begin with a quick review of the effects of small phase fluctuations in a non-topological junction.
The junction consists of a weak link between two superconductors with a junction capacitance $C$ described by the Hamiltonian
\begin{equation}
\hat{H} = E_C \left(\hat{n} - n_g \right)^2 - E_J \cos \hat{\theta},
\label{eqn:JJHamiltonian}
\end{equation}
where $E_J$ is the Josephson energy associated with the tunneling of Cooper pairs between the two superconductors, $E_C=\mathrm{e}^2/(2C)$, the charging energy of the weak link and $n_g$ the offset charge. 
The operator $\hat{n}$ measures the charge and the operator $\hat{\theta}$, measures the phase difference between the superconductors. 
To simplify the comparison with the following sections, we measure $\hat{n}$ (and $n_g$) in units of the electron charge $\mathrm{e}$, rather than in the more conventional units of $2\mathrm{e}$.
The commutation relation, $[\hat{\theta},\hat{n}]=2i$, therefore follows.
Several examples of superconducting circuits, such as the ones used in the Cooper pair box~\cite{1402-4896-1998-T76-024,RevModPhys.73.357}, quantronium~\cite{Vion886} and transmon~\cite{PhysRevA.76.042319} qubits, can be mapped to Eq.~\ref{eqn:JJHamiltonian}. In these circuits, $n_g$ is tuned using gate voltages, while the ratio of $E_J$ and $E_C$ may be tuned using split junctions or by adding additional capacitances (see e.g. Ref.~\onlinecite{CircuitQED}).

In the basis of phase eigenstates, the wave-function $\Psi(\theta) = \left\langle \theta | \Psi \right\rangle$ describing the Josephson junction follows the equation 
\begin{subequations}
\begin{equation}
\left[ E_C\left( -2 i \frac{d}{d \theta} -n_g \right)^2 - E_J \cos \theta \right]   \Psi (\theta) = E  \Psi (\theta) 
\end{equation}
respecting the boundary condition
\begin{equation}
\Psi (\theta + 2\pi) = \Psi(\theta).
\end{equation}
\end{subequations}
The dependence of the system on the offset charge $n_g$ can be transferred from the Schr\"{o}dinger's equation to the boundary condition via $\Psi(\theta)\rightarrow e^{i n_g \theta /2} \Psi( \theta)$ which results in,
\begin{subequations}
\begin{equation}
\left[ E_C\left( -2 i \frac{d}{d \theta} \right)^2 - E_J \cos \theta \right]   \Psi (\theta) = E  \Psi (\theta),
\label{eqn:JJeqn}
\end{equation}
\begin{equation}
\Psi (\theta + 2\pi) = e^{i \pi n_g}\Psi(\theta).
\label{eqn:JJbc}
\end{equation}
\label{eqn:JJtheta}
\end{subequations}
The above equations can be solved using Mathieu functions.
Nonetheless, expansions for different parameter regimes have been developed to provide more intuition.
Since we are interested in studying phase fluctuations, we focus on the $E_C \ll E_J$ limit.
This corresponds to the regime of interest of transmon qubits~\cite{PhysRevA.76.042319}. 

When $E_C \ll E_J $ the potential energy $- E_J \cos \theta$ dominates the energy of the system. 
Around the potential energy minima, at $\theta = 2\pi j$ with integer $j$, equation~\ref{eqn:JJeqn} can be mapped onto an harmonic oscillator having frequency $\hbar \omega = \sqrt{8 E_J E_C}$.
The low energy levels of the Josephson junction therefore correspond to harmonic oscillator levels. 
Deep inside the potential well, these harmonic oscillations do not depend on the boundary conditions given by Eq.~\ref{eqn:JJbc}.
To find the junction dependence on $n_g$, we need to account for quantum tunneling between the different potential minima.

Denoting the amplitude for quantum tunneling between the $m$th harmonic oscillator level of one of the potential minima and its nearest neighbors by $\nu_{m}$, it is possible to write an effective tight-binding Hamiltonian for the junction:

\begin{equation}
\hat{H} = \sum_{m=0}^\infty \sum_j \left[ \epsilon_m a_{m,j}^\dagger a_{m,j} -\nu_m  a_{m,j+1}^\dagger a_{m,j} + \text{h.c.}\right]
\label{eqn:TBH}
\end{equation} 
Here $a_{m,j}^\dagger$ is the creation operator for $m$th level of an harmonic oscillator around $2\pi j$, and $\epsilon_m =\hbar \omega(m+1/2) $ the energy of the a level. The tight-binding Hamiltonian in Eq.~\ref{eqn:TBH} is diagonalized using the operators $a_{m,k} = \sum_j e^{-ikj} a_{m,j}$:
\begin{equation}
\hat{H} = \sum_m \sum_k \left( \epsilon_m  - 2 \nu_m \cos k \right) a_{m,k}^\dagger a_{m,k}.
\end{equation} 
Comparing with Eq.~\ref{eqn:JJbc} leads to the identification $k=\pi n_g$, which allows us to conclude that for $E_C \ll E_J $ the dispersion of the $m$th level of the junction is given by
\begin{equation}
E_{m} (n_g) = \epsilon_m - 2 \nu_m \cos \left( \pi n_g \right), 
\end{equation}
which holds when $\nu_m \ll \hbar \omega$.

The tunneling amplitudes $\nu_m$ can be calculated using semi-classical methods. Here we briefly outline the calculation for the lowest energy level corresponding to the the phase slip probability $\nu_0$.  We use the dilute instanton gas approximation in the path integral imaginary time formalism (see e.g. Ref.~\onlinecite{doi:10.1142/9789812562197_0017}). In this formalism, the amplitude to propagate from $0$ to $2\pi$ during an imaginary time interval of length $2L$ is written as a weighted sum over all the paths that start at $0$ at time $\tau=-L$ and end at $2\pi$ at $\tau=L$:
\begin{equation}
(0, -L | 2\pi, L) = \int [\mathcal{D} \theta]  e^{-\frac{1}{\hbar} \int_{-L}^{L} \mathcal{L} \left(\theta(\tau) \right)d\tau},
\label{eqn:2pipaths}
\end{equation}
where 
\begin{equation}
\mathcal{L} \left( \theta \right) = \frac{\hbar^2 \left( \dtau \theta \right)^2}{16 E_C}+ E_J \left( 1 - \cos \theta \right)
\end{equation}
is commonly known as the sine-Gordon Lagrangian which is related to the Hamiltonian in Eq.~\ref{eqn:JJHamiltonian} through a Legendre transform.

For $L\rightarrow \infty$ the classical solution is a $2\pi$-kink also referred to as an instanton.  It is given by $\theta^{cl}_{2\pi} \left( \tau \right) = 4 \arctan \left( e^{\omega (\tau-\tau_0)} \right)$ where $\omega = \sqrt{E_J E_C}/\hbar$ coincides with the frequency of harmonic oscillations around the $2\pi j$ minima. Conversely, the model also has a classical solution with $\theta(-\infty)=2\pi$ and $\theta(\infty)=0$ known as an anti-kink.  
In the dilute instanton gas approximation, the path integration of Eq.~\ref{eqn:2pipaths} is done over combinations of kinks and anti-kinks and Gaussian fluctuations around them. 
Furthermore, it is assumed that the kinks and anti-kinks are separated enough (in imaginary time) that the interactions between them are negligible.
This yields the result
\begin{equation}
 \nu_0=\sqrt{2(\hbar \omega)^3/(\pi E_C)}e^{-\hbar \omega/E_C},
 \label{eqn:nu2pi}
\end{equation}
where $\hbar^2 \omega/E_C= \hbar \sqrt{8E_J/E_C}$ is the action of a $2\pi$ kink. 

To test the validity of Eq.~\ref{eqn:nu2pi} we ask whether the gas of kinks and anti-kinks is in fact dilute. This can be done by comparing the width of the kinks, $2/\omega$, with the expected average separation among them, $ \hbar/\nu_0$. The gas is dilute, and Eq.~\ref{eqn:nu2pi} is self-consistent, as long as $\nu_0 \ll \hbar \omega/2$, which is satisfied for $E_J \gg E_C$. 

This formalism can be extended to calculate the $n_g$-dependence of higher levels through the use of periodic instantons (see e.g.~Ref. \onlinecite{doi:10.1142/9789814397759_0026}). The decision to focus on $\nu_0$ was made for the sake of simplicity.

\section{Phase fluctuations in a topological Josephson junction}
\label{sec:Effects}

In a topological junction, the two superconductors coupled by the junction each present a Majorana mode close to the the junction. We denote these by $\gamma_1$ and $\gamma_2$, and ignore the other two Majorana modes which are far from the junction. The coupling of these Majorana modes adds a $4\pi$ periodic term to the tunneling current~\cite{1063-7869-44-10S-S29,Kwon2004,PhysRevB.79.161408,:/content/aip/journal/ltp/30/7/10.1063/1.1789931,PhysRevLett.105.077001,PhysRevLett.105.177002,Badiane2013840}.
The topological junction can then be modeled by the following Hamiltonian:
\begin{equation}
\hat{H} = E_C \left(\hat{n} - n_g \right)^2 - E_J \cos \hat{\theta} - i \gamma_1 \gamma_2 \frac{E_M}{2} \cos \frac{\hat{\theta}}{2}
\label{eqn:TJJ}
\end{equation}
where $i\gamma_1 \gamma_2$ is the parity of the fermionic mode caused by the hybridization the Majorana modes on both sides of the junction.

A physical realization of the above phenomenological model is possible using a Majorana Cooper pair box, such as the one studied in Ref.~\onlinecite{PhysRevB.90.075408}. In order to achieve the desired phase dominated limit the Majorana Cooper pair box could be shunted by a larger capacitance, as is done in transmon qubits~\cite{PhysRevA.76.042319}. 

If the local parity is conserved, the operator $i\gamma_1 \gamma_2$ in Eq.~\ref{eqn:TJJ} can be substituted by either one of its two eigenvalues $\pm 1$. Without loss of generality, from now on we assume $i\gamma_1 \gamma_2=1$. As long as the local parity is conserved, our results do not rest on this assumption. As in the previous section, after a charge translation the wave-function in phase basis follows an $n_g$ independent Schr\"{o}dinger's equation
\begin{subequations}
\begin{equation}
\left[ E_C\left(\! -2 i \frac{d}{d \theta} \right)^2\!  - E_J \cos \theta -\frac{E_M}{2} \cos \frac{\theta}{2} \right]  \Psi  = E  \Psi 
\label{eqn:TJJeqn}
\end{equation}
and a boundary condition 
\begin{equation}
\Psi (\theta + 4\pi) = e^{i 2 \pi n_g}\Psi(\theta).
\label{eqn:TJJbc}
\end{equation}
\label{eqn:TJJtheta}
\end{subequations}

As in the non-topological case, when $E_C$ is small compared to the tunneling $E_J$ and $E_M$ the energy is dominated by the tunneling terms which we refer to as ``the potential". We therefore expect the ground state wave function to be concentrated around the potential minima.  In the topological Josephson junction, the competition between the pair and single particle tunneling creates two different regimes depending whether the junction potential has a single minimum or a two minima for $0 \leq \theta < 4\pi$. 

\begin{figure}
\subfigure[$E_C=0.001$]{
\includegraphics[width=0.48\linewidth]{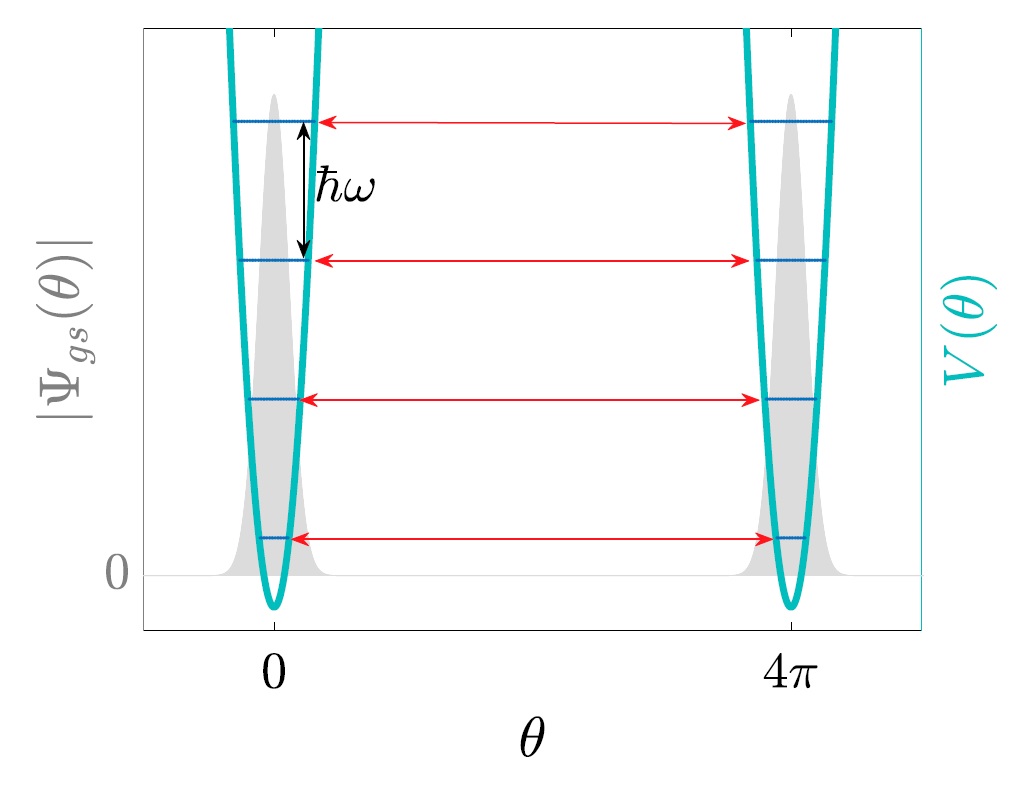}\label{fig:1minECsmall}}
\subfigure[$E_C=0.1$]{
\includegraphics[width=0.48\linewidth]{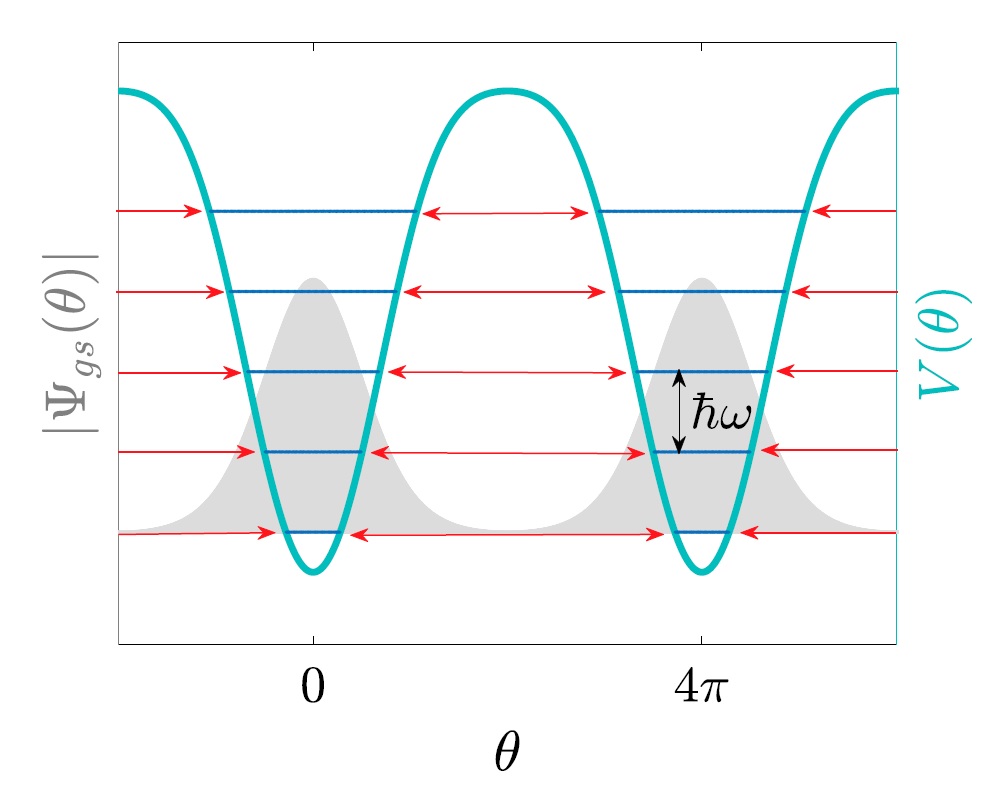}\label{fig:1minEClarge}}
\subfigure[$E_C=0.001$]{
\includegraphics[width=0.48\linewidth]{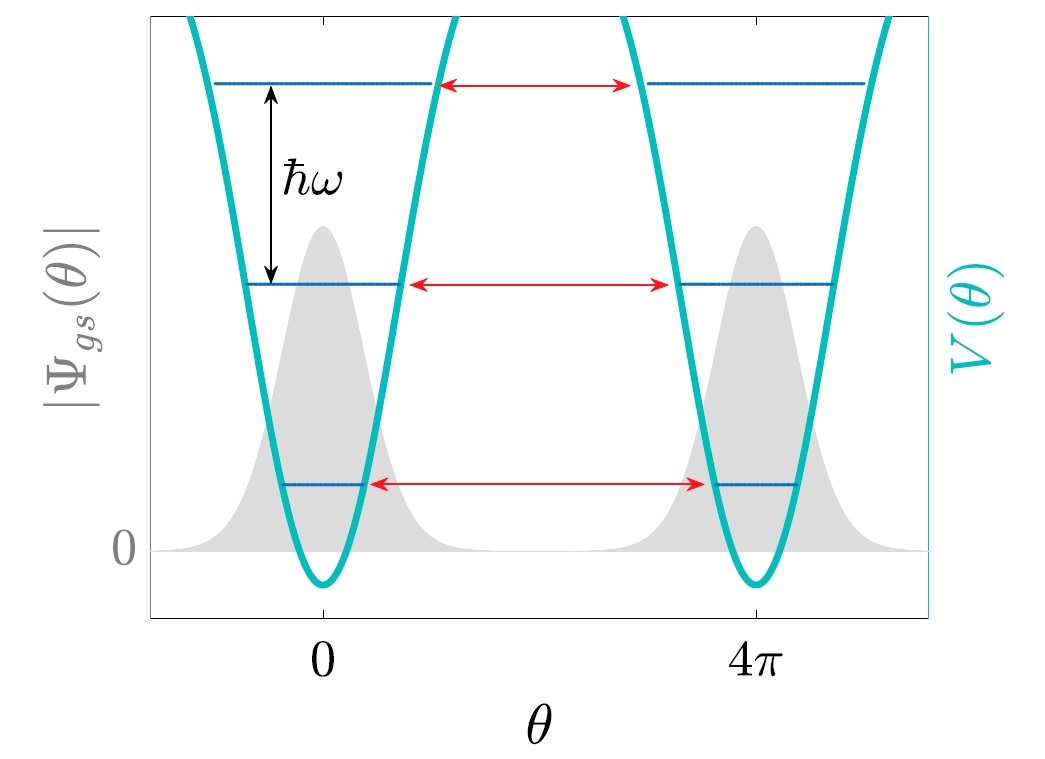}\label{fig:2minECsmall}}
\subfigure[$E_C=0.1$]{
\includegraphics[width=0.48\linewidth]{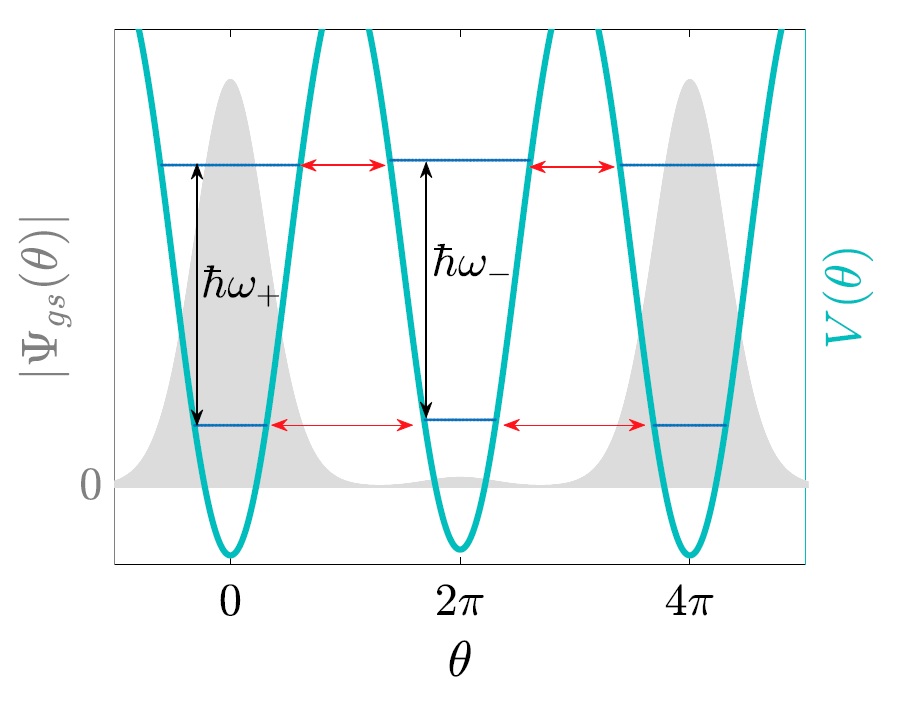}\label{fig:2minEClarge}}
\caption{Phase fluctuations in the single-minimum and double-minimum regimes of a topological Josephson junction. The first harmonic levels (blue lines) and the junction potential (green line) are shown for a junction with $E_M=2=10 E_J$ and (a) $E_C=0.001$ and (b) $E_C=0.1$, and a junction with $E_J=1=50 E_M$ and (c) $E_C=0.001$ and (d) $E_C=0.1$. The ground-state wave-functions (amplitudes shown in grey) correspond to linear superpositions of harmonic oscillations around the potential minima. The tunneling processes that give rise to the $n_g$ dispersion of each level are shown in red. In the double-minimum regime ((c) and (d)), increasing $E_C$ can change which are the dominant tunneling processes. The ground-state wave-function in (d) shows an additional (small) peak around $2\pi$. Note that in panels (c-d) the $2\pi$ minimum is not degenerate with the $4\pi$ ones.}
\label{fig:min}
\end{figure}

When $E_M/(8E_J)>1$, the junction potential has a single minimum in the $[0,4\pi)$ interval. Hence, the potential is minimized when $\theta = 4\pi m $ with $m$ an integer, and all the minima are degenerate. The frequency of harmonic oscillations around these minima, obtained by expanding Eq.~\ref{eqn:TJJeqn} around these values, is $\hbar \omega = \sqrt{8E_J E_C + E_M E_C}$. 
This is exemplified in Fig.~\ref{fig:1minECsmall} where the first few harmonic oscillator levels and the ground-state wave-function amplitude are shown for $E_M=2=10 E_J$ and $E_C=0.001$. 
The junction potential and the tunneling processes between the degenerate levels are also shown in Fig.~\ref{fig:1minECsmall}.
As $E_C$ increases, the spacing between the levels and tunneling amplitude increases and the harmonic wave-functions widen, as shown in Fig.~\ref{fig:1minEClarge} for $E_M=2=10 E_J$ and $E_C=0.1$.
However, the tunneling processes that give rise to the $n_g$ dispersion remain unchanged by the increase of $E_C$. 
In this regime, the topological junction behaves qualitatively similar to the non-topological junction from the previous section with half the $n_g$ periodicity and $4\pi$ phase slips taking the role of $2\pi$ phase slips. 

On the other hand, if $E_M/(8E_J)<1$, the junction potential has two minima in the $[0,4\pi)$ interval. 
Hence, the potential has two kinds of minima with two different frequencies for harmonic oscillations around them: $\theta = 4\pi m $ with frequency $\hbar \omega_+ = \sqrt{8E_J E_C + E_M E_C}$, and $\theta = 4\pi m + 2\pi $ with frequency $\hbar \omega_- = \sqrt{8E_J E_C - E_M E_C}$.  
In addition to the effects discussed in the previous paragraph, changing $E_C$ may also change the tunneling processes that contribute to each energy level. 
This is shown in Figs.~\ref{fig:2minECsmall} and~\ref{fig:2minEClarge}. The ground-state wave-function in Fig.~\ref{fig:2minECsmall} is peaked around $0$ and $4\pi$, whereas the ground-state wave-function in Fig.~\ref{fig:2minEClarge} shows additional contributions from oscillations around $2\pi$. 

\section{Effective models}
\label{sec:Effective}
In this section, we will discuss two different effective models for the junction ground-state: one in which only oscillations between $4\pi m$ minima contribute and one in which oscillations around all $2\pi m$ minima contribute to the ground-state.
We calculate the effective hopping parameters of each model and discuss their regions of validity. 

\subsection{$4\pi$ QPS model}

We can write an effective Hamiltonian for the ground-state of the junction as a combination of harmonic oscillations around $4\pi j$ plus hopping between such minima:
\begin{equation}
\hat{H} = \sum_j \left( \frac{\hbar \omega}{2} a_{j}^\dagger a_{j}   -\nu_{4\pi}  a_{j+1}^\dagger a_{j}  + \text{h.c.}\right),
\label{eqn:TBHT}
\end{equation} 
where $ \omega$ the frequency of harmonic oscillations around the minima at $4\pi j$ and is given by $\hbar \omega=\sqrt{8 E_J E_C + E_M E_C}$.
Accounting for the boundary condition, Eq.~\ref{eqn:TJJbc}, results in the following ground-state energy dispersion
\begin{equation}
E_{gs} (n_g) = \frac{\hbar \omega}{2}- 2 \nu_{4\pi} \cos \left(2 \pi n_g \right). 
\label{eqn:4piPSdispersion}
\end{equation}
This model gives an effective description of the system in the single-minimum regime and in the double-minimum regime for small enough $E_C$ (see Fig.~\ref{fig:min}).

The tunneling amplitude $\nu_{4\pi}$ can be calculated following the procedure outlined in Sec.~\ref{sec:review}. The imaginary time Lagrangian of the topological junction,
\begin{equation}
\begin{split}
\mathcal{L} \left( \theta \right) = &\frac{\hbar^2 \!\left( \partial_\tau \theta \right)^2}{16 E_C}\! +\! E_J \left( \!1 - \!\cos \theta \right) +\! \frac{E_M}{2}\! \left( \!1-\! \cos\frac{\theta}{2}\!\right)\!,\!
\end{split}
\label{eqn:DSGL}
\end{equation}
is known as the double sine-Gordon Lagrangian and its semi-classical dynamics have been widely studied.~\cite{Mussardo2004189}

\begin{figure}
	\includegraphics[width = 0.8\columnwidth]{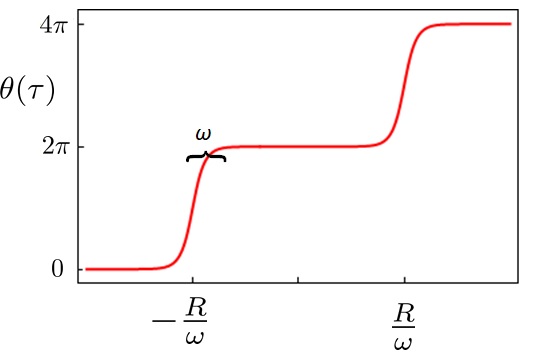}
	\caption{A $4\pi$ instanton is made of two $2\pi$ instantons of width $\omega$ separated by an imaginary time of $2R/\omega$.}
	\label{fig:double_instanton}
\end{figure}

Interestingly, as shown in Fig.~\ref{fig:double_instanton}, the $4\pi$ kink in the DSG model can be written as a sum of two $2\pi$ SG kinks:
\begin{subequations} 
\begin{equation}
\theta^{cl}_{4\pi} = 4 \arctan e^{\omega \left(\tau-\tau_0\right)-R} + 4 \arctan  e^{\omega \left( \tau-\tau_0 \right)+ R }. 
\end{equation}
The imaginary time separation of the two $2\pi$ kinks, $2R/\omega$, is set by the ratio of $E_M$ and $8 E_J$ as $R$ is given by:
\begin{equation}
R = \arccosh \left( \sqrt{1 + \frac{8E_J}{E_M}} \right).
\label{eqn:R}
\end{equation}
\end{subequations}
The $4\pi$ DSG kink is depicted in Fig.~\ref{fig:double_instanton}. The width of the kinks is controlled by the Josephson tunneling and the capacitance energy through $\omega = \sqrt{E_J E_C}/\hbar$ and the separation between kinks is controlled by the ratio of $E_M$ and $E_J$ through $2R/\omega$.
When $E_M \rightarrow 0 $, the separation between the two $2\pi$ kinks diverges ($R \rightarrow \infty$) meaning that the $4\pi$ kinks effectively decouple into two separate $2\pi$ kinks as the DSG Lagrangian reduces to the SG Lagrangian. 

Using the dilute instanton gas approximation, as before, we find
\begin{subequations}\label{eqn:nu4pi_first}
\begin{equation}
\nu_{4\pi}= \sqrt{\frac{8(\hbar \omega)^5}{\pi E_M E_C^2}} \exp \left(
- \frac{\hbar\omega}{E_C} \times f \left(\frac{E_M}{8 E_J} \right) \right)
\end{equation} 
where
\begin{equation}
f \left(x \right) = 2 + \frac{2x}{\sqrt{1+x}}\coth^{-1} \left(\sqrt{1+x} \right)
\end{equation}
\label{eqn:nu4piharmonic}\end{subequations}
is an increasing function with $f(0)=2$ and $f(\infty)=4$. With the appropriate modifications, this result is in agreement with the result found by Ref.~\onlinecite{PhysRevB.26.226} in a of statistical mechanics context. A more detailed derivation of how Eq.~\ref{eqn:nu4piharmonic} is obtained is shown in Appendix~\ref{sec:nu4pi}.

When $E_M \rightarrow 0$, $\nu_{4\pi}$ presents a square root divergence, i.e.  $\nu_{4\pi} \sim 1/\sqrt{E_M}$. This divergence has two physical interpretations. First, it is indicative of a resonance in tunneling~\cite{Singh} when $E_M\rightarrow 0$. In our context, it is a sign that the validity of the model breaks down in this limit. Secondly, this divergence is indicative of the restoration of a symmetry. In this case, the symmetry that is restored is the $2\pi$ translation symmetry; i.e. the decoupling of the two $2\pi$ kinks.

The restoration of the $2\pi$ translation symmetry for $E_M \rightarrow 0$ diminishes the range of the validity of the calculated expression for $\nu_{4\pi}$.  
This can be seen by noting that the dilute instanton gas approximation breaks down when $E_M \rightarrow 0$: the width of the $4\pi$ kinks $(2+2 R)/\omega $ diverges as $ -\log E_M$ whereas the average separation between the kinks $\hbar/\nu_{4\pi}$ goes to zero as $\sqrt{E_M}$. The assumption that the width of the $4\pi$ kinks is much smaller than the average separation between the kinks fails for $E_M \rightarrow 0$. We address this problem in the next subsection.

\subsubsection*{Emergent translational mode correction}

In order to derive a semiclassical expression whose validity extends to smaller $E_M/E_J$ ratios we account for a higher order of fluctuations in the direction of the emergent translational mode.~\cite{PhysRevB.35.3496,PhysRevB.40.686} Since the emergent translational mode is related to the decoupling of the two kinks, this is roughly equivalent to letting the distance between the two kinks fluctuate around its equilibrium value, $2R/\omega$. 
 
The result of Ref.~\onlinecite{PhysRevB.40.686} can be written in terms of $R$ as:
\begin{subequations}
\begin{equation}
\nu_{4\pi} = \frac{4 F(R)\left( \hbar \omega \right)^2}{\pi E_C} \:  \mathcal{I} \left(R,\tfrac{\hbar \omega}{E_C}\right) 
\label{eqn:nu4pi}
\end{equation}
where $F(R)$ is a numerical factor bounded by $\sqrt{2/5} \leq F(R) \leq 1$ and given by
\begin{equation}
F(R) = \frac{\sqrt{\cosh 2 R-R \tanh R-3 R \coth R+2}}{\sinh R \sqrt{2-8 R^2 \text{csch}^2 2 R }};
\end{equation}
and
\begin{equation}
\mathcal{I}\left(R, \alpha \right)  = \int_{0}^{\infty} dr \sqrt{1-4 r^2 \text{csch}^2(2 r)} e^{- \alpha S_R (r)}
\label{eqn:I}
\end{equation}
with
\begin{equation}
\begin{split}
S_R (r) =&   1+ \frac{\tanh^2 R}{\tanh^2 r} + 2r  \times \\
&\left( \frac{1}{\sinh 2r} + \frac{\coth r}{\cosh^2 R} - \frac{\tanh^2 R \coth r}{2 \sinh^2 r} \right).\\
\end{split}
\end{equation}\label{eqn:nu4piall}\end{subequations}
In the above expressions, $2r/\omega$ corresponds to the fluctuating distance between the two kinks and $S_R(r)$ is an $r$-dependent effective action which is minimized at $r=R$. For more details on how this expression is obtained, see Appendix~\ref{sec:RApp} and Ref.~\onlinecite{PhysRevB.40.686}.

To the best of our knowledge, a closed form expression for $\mathcal{I}\left(R, \alpha \right) $ does not exist.
Nonetheless, we can find approximate expressions for $\mathcal{I}\left(R, \alpha \right) $ for small and large $R$.
For small $R$, the greatest contribution to $\mathcal{I}\left(R, \alpha \right)$ comes from the $r$ values around $R$. 
A saddle point approximation of the integral $\mathcal{I}\left(R, \alpha \right)$ results in 
\begin{equation}
\begin{split}
\mathcal{I}\left(R, \alpha \right) & \approx  \sqrt{\frac{\pi}{2\alpha}} \frac{\cosh R}{F(R)}e^{- \alpha S_R(R)} .
\label{eqn:saddleaprox}
\end{split}
\end{equation}
This is a good approximation to $\mathcal{I}\left(R, \alpha \right)$ if $e^{2R} \ll 16 \alpha$ (see Appendix~\ref{sec:RApp}). 
Substituting this in Eq.~\ref{eqn:nu4pi} gives the expression for $\nu_{4\pi}$ obtained without including corrections due to the emergent translational mode, i.e. Eq.~\ref{eqn:nu4piharmonic}. Hence, Eq.~\ref{eqn:nu4piharmonic} is valid when $E_M/(8E_J) \gg  E_C/(4 \hbar \omega)$. 

When $R$ is large, the integral is dominated by the linear large $r$ behavior of $S_R(r)$. In Appendix~\ref{sec:largeR}, we find that for $16 \alpha^2 \ll e^{2R}$
\begin{equation}
\begin{split}
\mathcal{I}\left(R, \alpha \right)  \approx \frac{\cosh ^2(R) e^{-\alpha  \left(\tanh ^2(R)+1\right)}}{2 \alpha }. 
\end{split}
\label{eqn:largeRapprox}
\end{equation}
This leads to $\nu_{4\pi}\approx\nu_{4\pi}^{lr}$ with
\begin{equation}
\nu_{4\pi}^{lr} = \frac{f_2 \left( \frac{E_M}{8 E_J} \right) \left( \hbar \omega \right)^3 }{\pi E_C E_M}\exp \left[ - \frac{\hbar \omega}{E_C} \times f_1 \left( \frac{E_M}{8 E_J} \right) \right]
\label{eqn:4pismallEM}
\end{equation}
when $E_M/(8 E_J) \ll 0.25  E_C^2/(\hbar \omega)^2$. In the above equation, $f_1(x)$ and $f_2 (x)$ are order 1 numerical factors which decrease with $x$; their exact form can be found in Appendix~\ref{sec:largeR}.
Note that according to the above calculations $\nu_{4\pi}$ diverges for $E_M\rightarrow 0 $ as $1/E_M$. 

\subsection{Coupled $2\pi$ QPS model}

If the junction parameters are such that there are additional (local) minima at $2\pi m$ with $m$ odd and oscillations around those minima contribute to the ground-state (see e.g. Fig.~\ref{fig:2minEClarge}), we can describe it by the following effective Hamiltonian: 
\begin{equation}
\hat{H} = \sum_j \left(  \epsilon_j  a_j^\dagger a_j  -\nu_{2\pi}  a_{j+1}^\dagger a_j - \nu_{2\pi}  a_j^\dagger a_{j+1}\right),
\label{eqn:TBHsmallEM}
\end{equation} 
where $\nu_{2\pi} $ corresponds to tunneling amplitude between potential minima separated by $2\pi$. The energies $\epsilon_n$ are given by:
\begin{equation}
\begin{split}
\epsilon_{2n} = & \epsilon_{e} =   \frac{\hbar\omega_{+}}{2} \\
\epsilon_{2n+1} =& \epsilon_{o}=  E_M + \frac{\hbar \omega_{-}}{2} \\
\hbar \omega_\pm =& \sqrt{8E_J E_C\pm E_M E_C }. 
\end{split}
\end{equation} 
The dispersion of Eq.~\ref{eqn:TBHsmallEM} is 
\begin{equation}
\begin{split}
E_{\pm} (n_g) = &\frac{1}{2}(\epsilon_o + \epsilon_e) 
\\& \pm \frac{1}{2} \sqrt{(\epsilon_o -\epsilon_e )^2  + 8 \nu_{2\pi}^2 (1+\cos (2\pi n_g))} .
\end{split}
\label{eqn:2pidispersion}
\end{equation}

The hopping $\nu_{2\pi}$ can be calculated using the formula proposed by Ref.~\onlinecite{PhysRevA.86.012106} for the tunneling through an asymmetric potential. 
Without loss of generality, we can focus on calculating the amplitude for tunneling between $0$ and $2\pi$. The minimum at $0$ and the minimum at $2\pi$ are separated by a barrier which is largest at $\theta_{max}$. Following Ref.~\onlinecite{PhysRevA.86.012106} we define two potentials symmetric around $\theta_{max}$, $V_L(\theta)$ and $V_R(\theta)$, such that $V_L(\theta)$ ($V_R(\theta)$) is equal to the junction potential for $0<\theta<\theta_{max}$ ($\theta_{max}<\theta<2\pi$). Then $\nu_{2\pi}$ can be written as:
\begin{equation}\label{eqn:geometric_average}
\nu_{2\pi} = A \sqrt{\nu_L \nu_R},
\end{equation}
where $\nu_s$, $s=L,R$, is the probability for tunneling from $0$ to $2\pi$ through the potential $V_s$ and
\begin{equation}
A = \frac{1}{2} \left[\left( \frac{V_{max}-\epsilon_e}{V_{max}-\epsilon_o}\right)^{1/4}+\left( \frac{V_{max}-\epsilon_o}{V_{max}-\epsilon_e}\right)^{1/4} \right]^{1/2},
\end{equation}
with $V_{max}=V(\theta_{max})$. The above expression for $\nu_{2\pi}$ clearly breaks down when $\epsilon_o\geq V_{max}$; at that point the zero point motion of the shallow minimum becomes larger than the potential barrier. The approximations leading to the above expression for $\nu_{2\pi}$ start failing before this point.

For our model of a topological Josephson junction, $\theta_{max}$ and $V_{max}$ are given by:
\begin{equation}
\begin{split}
\theta_{max} = & 4 \arctan \left(\omega_+ /\omega_- \right) \\
V_{max} = & 2 E_J \left(E_M/(8E_J)+1\right)^2.
\end{split}
\end{equation}
And the $\theta_{max}$-symmetric potentials $V_L$ and $V_R$ are well approximated by
\begin{equation}
\begin{split}
V_{L} (\theta)& \approx E_J \left(1+\frac{E_M}{8E_J}\right)^2 \left( 1 - \cos \left( \frac{\pi \theta}{\theta_{max}}\right) \right) \\
V_{R} (\theta) &\approx E_M + \\& E_J \left(1-\frac{E_M}{8E_J}\right)^2 \left( 1 - \cos \left( \frac{\pi (\theta-2\pi)}{\theta_{max}-2\pi}\right) \right)\!,\!
\end{split}
\end{equation}
which leads to the following tunneling amplitudes: 
\begin{equation}\label{eqn:nu_s}
\nu_s = \frac{4}{\sqrt{P_s \pi }} \left( 8 E_{s}^3 E_C \right)^{1/4}
e^{-P_s \sqrt{\frac{8E_{s}}{E_C}} }
\end{equation}
with $P_L = \theta_{max}/\pi = 2 -P_R $, $E_L = E_J(1 + E_M/(8E_J))^2$ and $E_R = E_J (1-E_M/(8E_J))^2$. $P_s$ and $E_s$ are, respectively, the period and amplitude of the potential $V_s$ for $s=L,R$. 

For $E_M \rightarrow 0$ the dispersion in Eq.~\ref{eqn:2pidispersion} becomes 
\begin{equation}
E_{\pm} (n_g) \rightarrow  \frac{\hbar \omega}{2}  \pm \left| 2 \nu_0 \cos \left( \pi n_g \right) \right|. 
\label{eqn:EM0}
\end{equation}
This is the expected result for the $E_M\rightarrow0$ limit, as it corresponds to the breaking of the symmetry between the minima at even and odd multiples of $2\pi$ ``folding" the $n_g$-Brillouin zone. 

\begin{figure}
\centering
\includegraphics[width=0.8\linewidth]{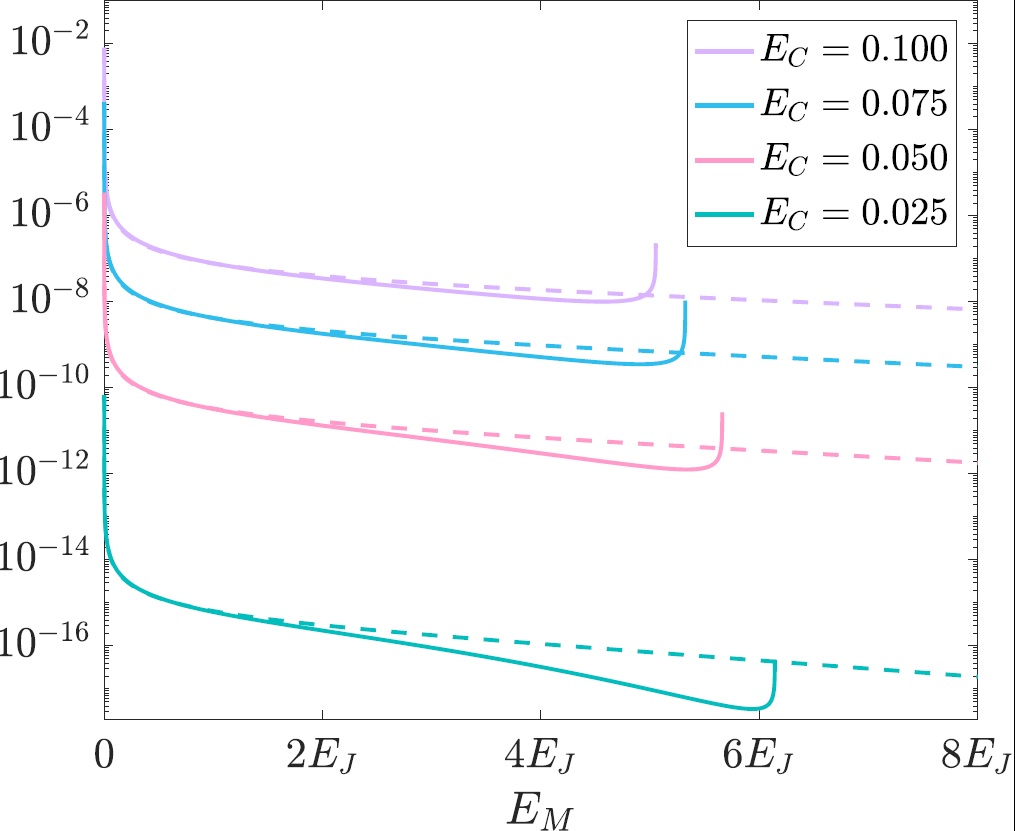}
\caption{Comparison of $\nu_{2\pi}^2 /|\epsilon_o -\epsilon_e| $ (solid line) and $\nu_{4\pi}^{lr}$ (dashed line) for $E_J=1$. The $\nu_{2\pi}^2 /|\epsilon_o -\epsilon_e| $ lines stop when the potential barrier is smaller than the zero point motion energy for oscillations around the shallow minima $\epsilon_o$.}
\label{fig:nu2pi4picompare}
\end{figure}

We also note that for $\nu_{2\pi} \ll |\epsilon_o -\epsilon_e |$ the lowest of the two bands becomes
\begin{equation}
E_{-} (n_g) \approx \epsilon_e - \frac{2 \nu_{2\pi}^2 }{|\epsilon_o -\epsilon_e|} -\frac{2 \nu_{2\pi}^2 }{|\epsilon_o -\epsilon_e|} \cos (2\pi n_g) .
\end{equation}
This dispersion would be equivalent to the dispersion found for the $4\pi$ phase slip model (\ref{eqn:4piPSdispersion}) if $\nu_{2\pi}^2 /|\epsilon_o -\epsilon_e| \rightarrow \nu_{4\pi}$. As shown in Fig.~\ref{fig:nu2pi4picompare}, we find that $\nu_{2\pi}^2 /|\epsilon_o -\epsilon_e| \approx \nu_{4\pi}^{lr}$. This allows us to interpret $\nu_{4\pi}^{lr}$ as arising from coupled but not confined $2\pi$ phase slips.

\subsection{Validity of the effective models}

\begin{figure}
\subfigure[$E_C=0.005$]{
\includegraphics[width=0.58\linewidth]{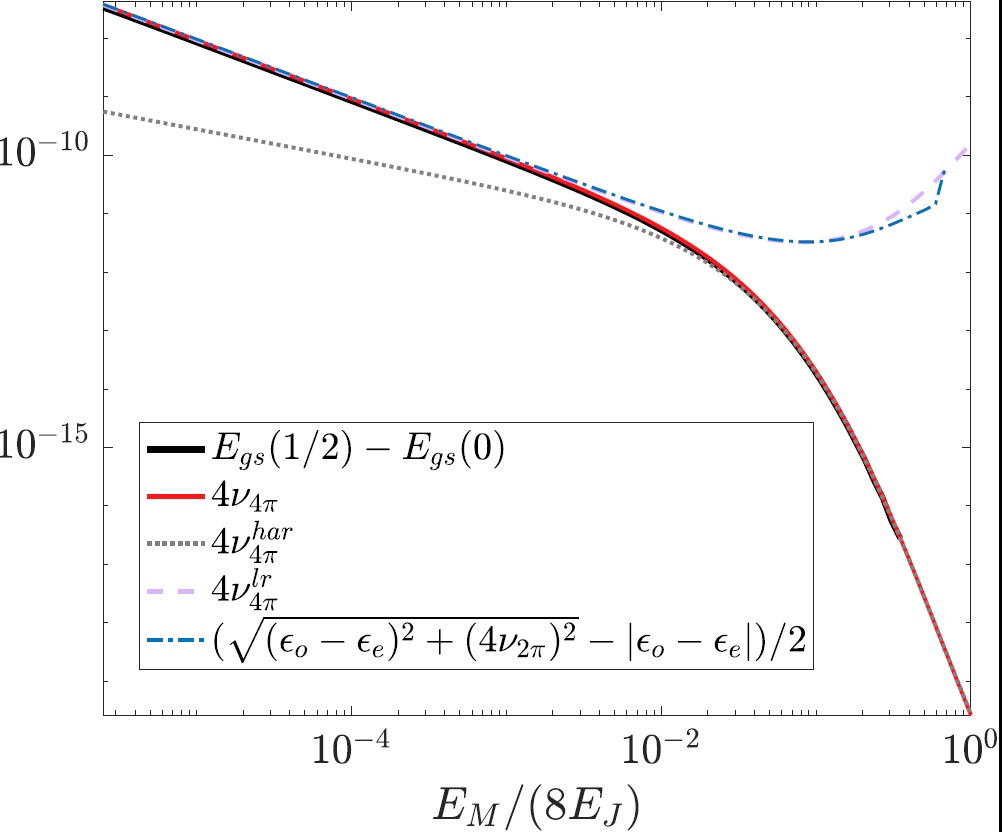}\label{fig:EC1}}
\subfigure[$E_C=0.01$]{
\includegraphics[width=0.58\linewidth]{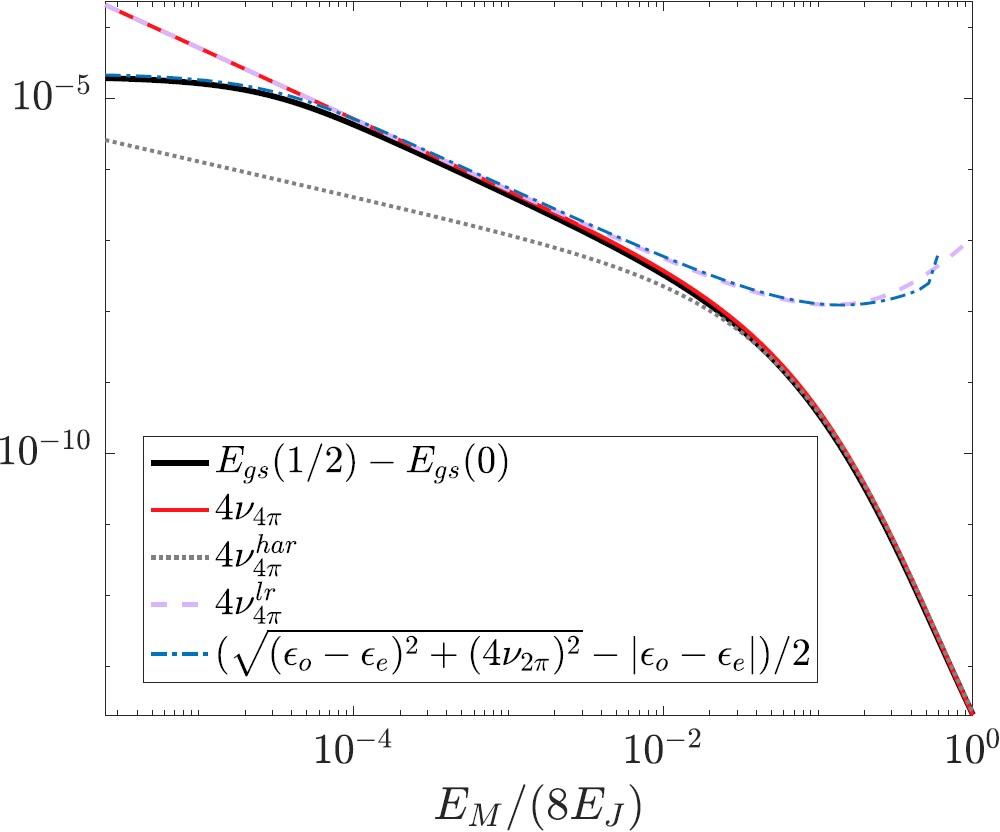}\label{fig:EC2}}
\subfigure[$E_C=0.02$.]{
\includegraphics[width=0.58\linewidth]{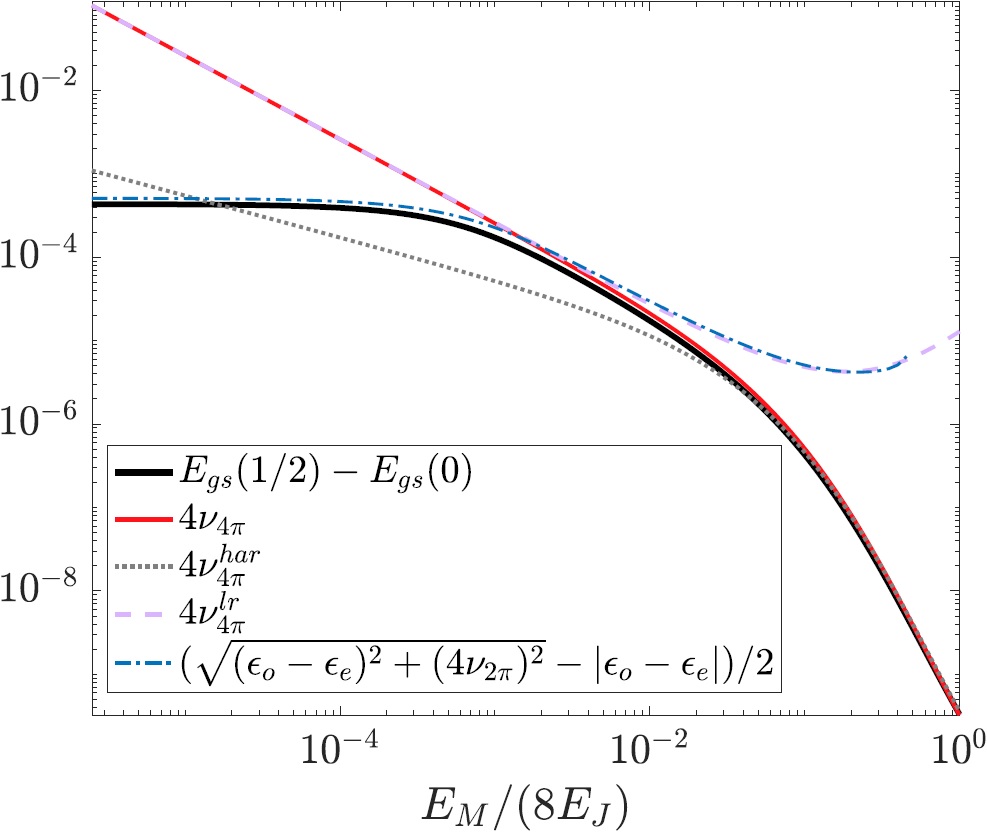}\label{fig:EC3}}
\subfigure[$E_C=0.05$.]{
\includegraphics[width=0.58\linewidth]{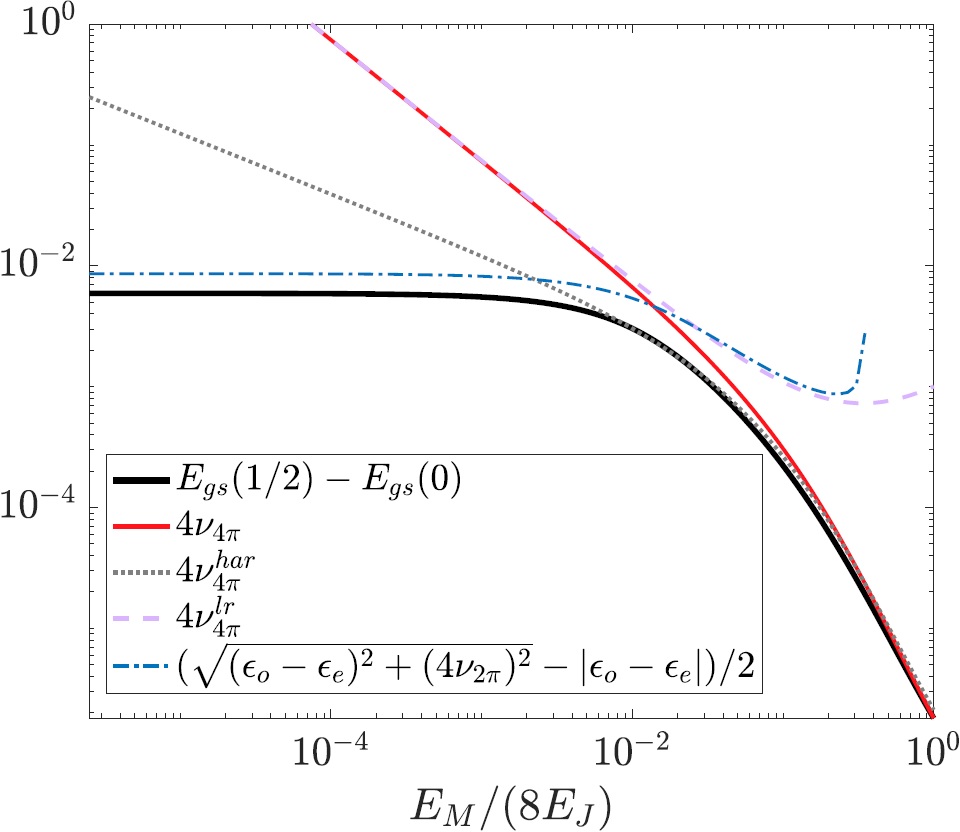}\label{fig:EC4}}
\caption{The ground state energy of the model as a function of the control parameter $n_g$ depends on the phase slip probability. We therefore use the quantity $E_{gs}(1/2)-E_{gs}(0)$ to benchmark the effective models against numerical results (black solid line). If only $4\pi$ phase slips are present  (Eq.~\ref{eqn:4piPSdispersion}), $E_{gs}(1/2)-E_{gs}(0)=4\nu_{4\pi}$. The different obtained expressions for $4\nu_{4\pi}$, Eqs.~\ref{eqn:nu4pi_first} (solid red line), Eqs.~\ref{eqn:nu4pi}-\ref{eqn:saddleaprox} (dotted gray) and Eq.~\ref{eqn:4pismallEM} (dashed purple), are shown. When $2\pi$ phase slips dominate (Eq.~\ref{eqn:2pidispersion}), $E_{gs}(1/2)-E_{gs}(0) = (\sqrt{(\epsilon_o-\epsilon_e)^2 + (4\nu_{2\pi})^2}-|\epsilon_o-\epsilon_e|)/2$ with $\nu_{2\pi}$ given by Eq.~\ref{eqn:nu2pi} (blue dashed-dotted). The graphs are shown as a function of $E_M/(8 E_J)$ with the sum $8E_J+E_M$ fixed at 1.}
\label{fig:Comparison}
\end{figure}

The image that emerges from the results in this section and the previous energetic considerations is as follows. Given a junction with fixed $E_M$ and $E_J$ values, it is always possible to find $E_C$ small enough such that the junction is well described by $4\pi$ QPS. On the other hand, given $E_C$ we can always find a $E_M$ small enough so $2\pi$ QPS are still present in the system. The range of $E_C$ values in which the junction can be fully described by $4\pi$ QPS processes shrinks to $0$ when $E_M\rightarrow 0$. To further clarify the range of parameters in which each picture is valid, we compare the different effective models for the topological Josephson junction with numerical result.

The spectrum of Eq.~\ref{eqn:TJJ} is obtained numerically by truncating the Hilbert space in the number basis, where the Hamiltonian becomes
\begin{equation}
\begin{split}
 H =  \sum_{n=-\infty}^\infty \left[ E_C \left(n-n_g \right)^2 \ket{n}\bra{n} - \tfrac{E_M}{4} (\ket{n}\bra{n+1} \right. \\
 + \left.\ket{n}\bra{n-1}) - \tfrac{E_J}{2} ( \ket{n}\bra{n+2}  + \ket{n}\bra{n-2}) \right].
\end{split}
\label{eqn:discreteH}
\end{equation}
The numerical results shown in this paper are obtained by taking the sum in the above equation from $n=-N$ to $n=N$ with $N=10^4$. 

Comparisons between $E_{gs} (n_g)$ for the topological Josephson junction predicted by the effective models discussed previously and numerical results are shown in Fig.~\ref{fig:Comparison}. The comparisons are done by plotting the difference $E_{gs}(1/2)-E_{gs}(0)$ as a function of $E_M/(8E_J)$ for different values of $E_C$. In Fig.~\ref{fig:Comparison} we have fixed $8E_J+E_M=1$ such that $\hbar \omega$ is kept constant throughout each plot; this is done to show the entire range of $E_M/(8E_J)$ in the same plot. 
As expected, when $E_M/(8E_J)\rightarrow 0$ the numerical results (solid black line) agree with the $2\pi$ QPS description (dotted-dashed blue line) provided by the tight-binding Hamiltonian, Eq.~\ref{eqn:TBHsmallEM}. While for larger values of $E_M/(8E_J)$ the $4\pi$ QPS description, i.e. that of Eq.~\ref{eqn:TBHT}, is closer to the numerical results. In addition, increasing $E_C$ reduces the range of $E_M/(8E_J)$ in which the $4\pi$ QPS description is valid. For instance, in Fig.~\ref{fig:EC1} the numerical results agree with the $4\pi$ QPS description for all the values $E_M/(8E_J)$ in the plot. While in Fig.~\ref{fig:EC2}, this only occurs for $E_M/(8E_J) \in (10^{-4},1)$ and; in Fig.~\ref{fig:EC3} and Fig.~\ref{fig:EC4} for $E_M/(8E_J) \in (10^{-2},1)$ and  $E_M/(8E_J) \in (0.1,1)$, respectively.  

Fig.~\ref{fig:Comparison} also shows the results of the $4\pi$ QPS description of the junction using the two approximations found for $\nu_{4\pi}$: the small $E_M$ approximation given by Eq.~\ref{eqn:4pismallEM} (purple dashed lines) and the large $E_M$ approximation of Eq.~\ref{eqn:nu4piharmonic} (gray dotted lines). Finally, Fig.~\ref{fig:Comparison} shows that as $E_C$ increases the QPS descriptions of the topological Josephson junction become less accurate. This is expected, as all the expressions for the tunneling amplitudes are obtained using the dilute instanton gas approximation which relies on $E_C \ll \hbar \omega$ and thus becomes less accurate as $E_C$ increases. 

We can use the numerical results to figure out the range of parameters in which each picture is more appropriate. This is shown in Fig.~\ref{fig:Main}. As discussed previously, close to the boundary between the coupled $2\pi$ QPS and the $4\pi$ QPS regions, both descriptions give similar results.

\section{Connection with the large charging energy limit}
The results above indicate that the presence of local minima at odd multiples of $2\pi$ in the junction potential lead to a ground-state wave-function weight at odd multiples of $2\pi$ if $E_C$ is large enough.
A question that arises is whether the presence of local minima in the potential guarantees that there will be a large enough $E_C$ such that the ground-state wave function is peaked at odd multiples of $2\pi$.
This can be answered by looking at the dominant charging energy limit.

For $E_M=E_J=0$, the eigenstates of the junction have a well defined particle number $n$ and their energies are given by $E_C(n-n_g)^2$. In the gauge where $\Psi(\theta+4\pi)=\Psi(\theta)$, the phase space wave-functions of such states are given by $\Psi(\theta)= e^{-il\theta/2}$ with integer $l$. If $E_M,E_J \ll E_C$ the eigenstates of the junction can be found perturbatively  from the well defined number states. To first order in perturbation theory, the ground-state of the junction for $n_g \in (-1/2,1/2)$ is given by the unnormalized wave function
\begin{equation}
\begin{split}
\Psi_{gs} (\theta) = &1 -\frac{E_M e^{-i\theta/2}}{4E_C(2n_g-1)}+\frac{E_Me^{i\theta/2}}{4E_C(2n_g+1)}\\& -\frac{E_J e^{-i\theta}}{8E_C(n_g-1)} +\frac{E_J e^{i\theta}}{8E_C(n_g+1)}.
\end{split}
\end{equation}
The above wave-function will be peaked at $2\pi$ if $|\Psi_{gs} (\theta)|$ has a local maximum at this point. 

For simplicity, we focus on $n_g=0$. In this case, 
\begin{equation}
\begin{split}
\Psi_{gs} (\theta) = 1 +\frac{E_M }{2E_C} \cos \frac{\theta}{2}+\frac{E_J }{4E_C} \cos \theta.
\end{split}
\end{equation}
Since $E_M \ll E_C$ and $E_J\ll E_C$ then $\Psi_{gs} (\theta) = |\Psi_{gs} (\theta)|$. Looking at the derivatives of $\Psi_{gs} $ at $\theta=2\pi$, we find that $2\pi$  is a minimum of $\Psi_{gs}$ when $E_M/(2 E_J) >1$ and a maximum when $E_M/(2 E_J) <1$. Therefore, the $n_g=0$ ground-state wave-function is peaked around odd multiples of $2\pi$ if $E_M/(2 E_J) <1$. 

We see then that for $2E_J<E_M<8E_J$ and $n_g=0$, the ground-state wave-function does not peak around odd multiples of $2\pi$ despite the junction having local potential minima there. Moreover, the wave-function weight around odd multiples of $2\pi$ is highest for integer values of $n_g$. This can be intuitively understood by noticing that the wave-function weight around odd multiples of $2\pi$ for half-integer $n_g$ is strongly suppressed as a result of the degeneracy between the two nearby $n$ states. Therefore, for any given $n_g$ the ground-state wave-function does not peak around odd multiples of $2\pi$ if $2E_J<E_M$. Then the presence of local minima in the potential does not guarantee that there will be a large enough value of $E_C$ to cause a ground-state wave-function peak at odd multiples of $2\pi$. 

\section{Discussion}
\label{sec:Dis}

In the above we found that for any ration of $E_M<8E_J$ one might find both $4\pi$ and $2\pi$ phase slips, depending on the strength of the phase fluctuations given by $E_C$. Therefore, the scenario of $4\pi$ phase slips only is bound to fail for some value of $E_C$. We have estimated the value of $E_C$ above which $2\pi$ phase slips dominate in the following way.  First we have evaluated the $4\pi$ phase slip probability $\nu_{4\pi}$ using a double sine-Gordon model. Well below the cross-over line in Fig.~\ref{fig:Main} a double instanton classical solution with gaussian quantum fluctuations yield Eq.~\ref{eqn:nu4pi_first}. This equation fails at low $E_M$ even before $2\pi$ phase slips take over due to a possible translational mode that was not taken into account. We improve the calculation in Eqs.~\ref{eqn:nu4piall} which does not have a closed form solution but may be approximated in the subsequent equations.In the $2\pi$ phase slip dominated regime we estimate $\nu_{2\pi}$ using a method for asymmetric barrier and arrive at Eqs.~\ref{eqn:geometric_average}-\ref{eqn:nu_s}. Using both the $2\pi$ and $4\pi$ phase slip scenario we generate plots for the energy difference $E_{gs}(n_g=1/2)-E_{gs}(n_g=0)$ which are compared with numerical solution for the problem in a truncated Hilbert space (using Eq.~\ref{eqn:discreteH}). The quality of the various approximations point to the cross-over depicted in Fig.~\ref{fig:Main}. A rough estimate of the cross-over as a function of the problem's energy scales $x=E_M/8E_J$ and $y=E_C/8E_J$ can be found by comparing $\nu_{4\pi}$ and $\nu_{2\pi}$ which yields $x\propto \exp\left(-\alpha/\sqrt{y} \right)$ with some slowly varying $\alpha(x)$. 

It is interesting to discuss the implications of our results on the dissipative transition that is expected in this system.~\cite{Panyukov1988,SCHON1990237,PhysRevLett.112.247001} This transition was previously studied in Ref.~\onlinecite{PhysRevLett.112.247001}, where it was found that the presence of $4\pi$ periodic tunneling would reduce the ohmic dissipation needed to restore superconductivity by a factor of 4. However, the results of Ref.~\onlinecite{PhysRevLett.112.247001} assumed that the topological junction could always be described by $4\pi$ QPS. In this work, we find that this is not necessarily the case. Consider a junction with fixed $E_J$ and $E_C$, when $E_M=0$ the junction is described by $2\pi$ QPS, turning on $E_M$ leads to an increasing coupling of this $2\pi$ QPS until they become confined into pairs. Following the critical dissipation throughout this same path would lead to a continuous decrease in it until it reaches $1/4$ of the original value at the point where the $2\pi$ QPS are fully suppressed. We also find that the critical dissipation needed to stabilize the superconductivity in our model of a topological Josephson junction is dependent on $E_C$. 

An important caveat of using the dissipative phase transition as a mechanism for detecting Majorana modes is that the dissipation induced by quasi-particle tunneling also reduces the critical resistance of non-topological Josephson junctions by a factor of 4. Furthermore, the effects of dissipation induced by quasi-particle tunneling in non-topological Josephson junctions are dependent on the ratio between the Josephson coupling and the charging energy.~\cite{SCHON1990237} This is because both the $4\pi$ periodic tunneling induced by Majoranas and the quasi-particle tunneling are single particle tunneling processes that break the same symmetry (the $2\pi$ periodicity of a non-topological Josephson junction), albeit the difference in coherence. A more careful analysis of dissipation in the topological Josephson junction is required to find whether there are signatures in the dissipative transition that would allow distinguishing between the $4\pi$ periodic tunneling induced by Majoranas and the quasi-particle tunneling.

The difference in the effects of $4\pi$ periodic vs. quasi-particle tunneling in the dissipative transition is unclear. However, the effects on the charge offset dispersion are clearly different. While both kinds of single particle tunneling turn the system from $2\pi$ periodic to $4\pi$ periodic, the Majorana assisted tunneling opens up a gap (see Eq.~\ref{eqn:2pidispersion}), while the quasi-particle tunneling does not~\cite{SCHON1990237}. This could be a potential probe to distinguish between the two kinds of single-particle tunneling. 

Finally, another important issue to consider is the effect of quasi-particle poisoning in this system. Since instanton techniques tend to be useful to describe systems coupled to external environments~\cite{RevModPhys.59.1}, the formalism used in this work could be useful to study the effects of quasi-particle poising.  

\section{Conclusions}
\label{sec:conclusion}

We studied the effects of phase fluctuations induced by charging effects in a simple model of a topological Josephson junction. Our model considers both single particle tunneling and pair tunneling, which are, respectively, $4\pi$ and $2\pi$ periodic with respect to the superconducting phase difference across the junction. We found that when the single particle tunneling is a small component of the total tunneling current there are two possible ways to describe the ground-state of the junction: 1) in terms of $4\pi$ QPS or 2) in terms of coupled $2\pi$ QPS. We found the tunneling amplitudes for both effective descriptions and compared them to numerical results to determine the range of parameter in which each description is appropriate. We note that in a real junction one may not have control over the relative strength of $E_M$ and $E_J$ and therefore observing $2\pi$ phase slips does not necessarily imply that the junction isn't topological. However, if the junction is indeed topological and phase fluctuations can be reduced through capacitance (reducing $E_C$), $2\pi$ phase slips will be suppressed revealing the topological nature of the junction.

In addition, we discussed the possible implications that our results have for the dissipative phase transition expected in this system. As was previously found by Ref.~\onlinecite{PhysRevLett.112.247001}, when the ground-state of the junction is described by $4\pi$ QPS we expect the critical resistance needed to make the junction superconducting to be 4 times smaller than the critical resistance needed to make a non-topological junction superconducting. In the regime where tunneling processes between minima separated by $2\pi$ are still present in the system, we expect the critical transition to be somewhere between these two critical values. Given that increasing the charging energy of the junction may change the tunneling processes present in the system, our results also point towards a charging energy dependence of the critical resistance for the dissipative transition.

Several questions regarding the dissipative transition, particularly in relation to quasi-particle tunneling, remain unanswered. In the future, we will use the formalism developed in this work to obtain a quantitative description of this transition. It would also be interesting to pinpoint the relation between the results presented in this work and the dominant charging energy limit. 

\section{Acknowlegments}
The authors thank D. Van Harlingen and S. Hegde for useful discussions. This material is based upon work supported by NSERC, FQRNT (RRM, TPB), the Secretary of Public Education and the Government of Mexico (RRM) and the National Science Foundation under Grant No. 1745304 (SV).

\appendix

\section{Path Integral Calculations}
\label{sec:ps}

\subsection{4$\pi$ phase slip amplitude}
\label{sec:nu4pi}

The calculation of the tunneling amplitude between the different potential minima can be performed using standard semi-classical methods. 
We include the calculation here in detail for completeness, largely following Ref.~[\onlinecite{doi:10.1142/9789812562197_0017}].

We begin by calculating the amplitude to propagate from $0$ to $4\pi$ in an imaginary time interval $2L$. This is given by the following path integral:
\begin{equation}
(0 , -L | 4\pi , L) = \int [\mathcal{D} \theta]  e^{-\frac{1}{\hbar} \int_{-L}^{L} \mathcal{L} \left(\theta(\tau) \right)d\tau},
\label{eqn:tun}
\end{equation}
where $\mathcal{L} (\tau)$ is the Double sine-Gordon (DSG) Lagrangian given by Eq.~\ref{eqn:DSGL}, which can be rewritten as,
\begin{equation}
\mathcal{L} \left( \theta \right) = M \left( \frac{\left( \partial_\tau \theta \right)^2}{2} + V(\theta)\right)
\end{equation}
with
\begin{equation}
\begin{split}
V(\theta)\!=\!\omega^2\!\left[\!\tanh^2\! R \left(\! 1\! -\! \cos\! \theta \right)\! +4\!\sech^2\! R\left( \!1 \!- \!\cos\!\frac{\theta}{2}\!\right) \!\right]
\end{split}
\label{eqn:potential}
\end{equation}
and 
\begin{equation}
\begin{split}
&M = \hbar^2/(8E_C) \\
 &\omega =\sqrt{E_C \left(8E_J + E_M \right)}/\hbar \\
&\cosh \left(R \right) =  \sqrt{(8E_J+E_M)/E_M}.
\end{split}
\label{eqn:massetc}
\end{equation}

We expect the leading contribution to the path integral to be from paths of the form
\begin{equation}
\theta\left( \tau \right) = \theta^{cl} \left( \tau\right) +\chi(\tau)
\label{eqn:paths}
\end{equation}
where $\theta^{cl} \left( \tau\right)$ is the classical path which minimizes the action and interpolates between $\theta = 0$ at $\tau=-L$ and $\theta=4\pi$ at $\tau= L$. $\chi(\pm L) = 0$ is the deviation from the classical path and $\theta^{cl} \left( \tau\right)$ fulfills the following equation:
\begin{equation}
\frac{dV}{d\theta} \left( \theta^{cl} \left( \tau\right)\right) = \frac{d^2\theta^{cl}}{d \tau^2} .
\label{eqn:classical}
\end{equation}
In the limit $L\rightarrow \infty$, $\theta^{cl} \left( \tau\right)$ is given by
\begin{equation}
\begin{split}
\theta^{cl} = 4\arctan[e^{\omega\left(\tau -\tau_0 \right)+R}]+ 4\arctan[e^{\omega\left(\tau -\tau_0 \right)-R}].
\end{split}
\label{eqn:kink}
\end{equation}

Up to second order in $\chi(\tau)$ the Lagrangian for paths of the form Eq.~\ref{eqn:paths} is
\begin{equation}
\begin{split}
\mathcal{L} \left( \theta \right) =& \mathcal{L} \left( \theta^{cl} \right) + \frac{M}{2}  \left(\partial_\tau \chi \right) ^2 + \frac{M}{2} \frac{d^2V}{d\theta^2}  \left( \theta^{cl} \right) \chi^2 \\& + M \partial_\tau ( \chi \partial_\tau \theta^{cl}) . 
\end{split}
\end{equation}
This allows us to split the path integral in Eq.~\ref{eqn:tun} into two parts:
\begin{equation}
(0 , -L | 4\pi , L)  \approx F \exp \left( -\frac{S^{cl}}{\hbar} \right)
\end{equation}
with $S^{cl}$ the action of the instanton, 
\begin{equation}
S^{cl} = \int_{-L}^{L} d\tau \mathcal{L} \left( \theta^{cl} \right) 
\end{equation}
and $F$ contains the sum over Gaussian fluctuations around such instanton. $F$ can be written as
\begin{equation}
F = \int [\mathcal{D} \chi] \exp \left(  -\frac{M}{2 \hbar}\int_{-L}^{L} d\tau  \chi D  \chi  \right),
\label{eqn:F}
\end{equation}
with $D$ is the following differential operator:
\begin{equation}
D = - \frac{d^2}{d \tau^2} + \frac{d^2 V}{d \theta^2} \left( \theta^{cl} \left( \tau \right)\right).
\label{eqn:D}
\end{equation}

The path integral in Eq.~\ref{eqn:F} can be solved expanding $\chi$ in terms of the eigenfunctions of the operator $D$, i.e. taking
\begin{equation}
\chi\left(\tau \right) = \sum_n \chi_n y_n\left(\tau \right)
\end{equation}
with
\begin{equation}
D y_n \left( \tau \right) = \lambda_n y_n.
\end{equation}
This leads to 
\begin{equation}
F = \mathcal{N} \prod_n\! \int_{-\infty}^{\infty}\! \frac{d \chi_n}{\sqrt{2 \pi \hbar /M}}  e^{-\frac{M \lambda_n \chi_n^2}{2 \hbar}}
\label{eqn:Gaussian}
\end{equation}
with $\mathcal{N}$ a normalization constant. However, the above expression in not well defined since the operator $D$ contains a zero mode, $\lambda_0$, which leads to a divergence in $F$. The time $\tau_0$ at which the kink solution is centered is arbitrary which leads to $D\partial_\tau \theta^{cl}=0$; i.e. the zero mode is a consequence of the time-translational invariance of the system.  To deal with this divergence, we use the Fadeev-Popov method to transform the $\chi_0$ integration to a $\tau_0$ integration. 

The Fadeev-Popov method consists of inserting
\begin{equation}
1 = \int d\tau_0 \left| \frac{\partial \chi_0}{\partial \tau_0} (\chi_0 = 0)\right|\delta(\chi_0), 
\end{equation}
into the expression for $F$ given by Eq.~\ref{eqn:Gaussian}:
\begin{equation}
\begin{split}
F = &\mathcal{N} \prod_{n=1}^\infty\! \int_{-\infty}^{\infty}\! \frac{d \chi_n}{\sqrt{2 \pi \hbar /M}}  e^{-\frac{M \lambda_n \chi_n^2}{2 \hbar}} \times \\
 &\int d\tau_0 \left| \frac{\partial \chi_0}{\partial \tau_0} (\chi_0 = 0)\right|
\int \frac{d \chi_0}{\sqrt{2 \pi \hbar /M}} \delta(\chi_0)\\
= &\mathcal{N} \prod_{n=1}^\infty\! \int_{-\infty}^{\infty}\! \frac{d \chi_n}{\sqrt{2 \pi \hbar /M}}  e^{-\frac{M \lambda_n \chi_n^2}{2 \hbar}} \times \\
& \int \frac{d \tau_0}{\sqrt{2 \pi \hbar /M}} \left| \frac{\partial \chi_0}{\partial \tau_0} (\chi_0 = 0)\right|.
\end{split}
\end{equation}
 The Jacobian $\left| \frac{\partial \chi_0}{\partial \tau_0} (\chi_0 = 0)\right|$ can be found rewriting the path $\theta$ so that fluctuations in the direction of the zero mode are traded for an explicit $\tau_0$ dependence:
\begin{equation}
\begin{split}
\theta (\tau )= & \theta^{cl}  \left( \tau - \tau_0 \right) + \sum_{n=1}^{\infty} \chi_n y_n\left(\tau - \tau_0 \right).
\end{split}
\label{eqn:newpath}
\end{equation} 
Comparing the above expression for the path with that of Eq.~\ref{eqn:paths} leads to
\begin{equation}
\begin{split}
\chi_0 =& f(\tau_0) + \sum_{m=1}^{\infty} \xi_m r_{n} (\tau_0)
\end{split}
\end{equation}
with 
\begin{equation}
\begin{split}
&f(\tau_0)\! =\! \int\! d \tau\! \left(\theta^{cl}\!  \left(\tau\! -\! \tau_0\! \right)\! -\! \theta^{cl}\! (\tau)\right)\! y_0 \left(\tau \right) \\
&r_{m} (\tau_0)\! =\! \int \!d \tau  y_m \left(\!\tau\! -\! \tau_0 \right)  y_0\left(\tau \right).
\end{split}
\end{equation}
Furthermore, we note that the constraint $\chi_0=0$ corresponds to $\tau_0=0$ so we obtain:
\begin{equation}
\left| \frac{\partial \chi_0}{\partial \tau_0} (\chi_0\!=\!0)\right| = \left|f^\prime(0) + \sum_{m=1}^{\infty} \xi_m  r_{m}^\prime (0)\right| \\
\end{equation}

We know $\partial_\tau \theta^{cl} \propto y_0 \left( \tau \right)$ since $D\partial_\tau \theta^{cl}=0$. The proportionality constant can be found using the following expression:
\begin{equation}
\int_{-\infty}^{\infty} d \tau  (\partial_\tau \theta^{cl})^2 = \frac{S^{cl}}{M},
\end{equation}
which stems from the fact that $\theta^{cl}(\tau)$ minimizes the action (Eq.~\ref{eqn:classical}). 
We use this to find $f^\prime(0)$:
\begin{equation}
\begin{split}
f^\prime(0) &=\! -\! \int d \tau  \partial_\tau \theta^{cl} \! \left( \tau  \right)  y_0\!\left(\tau \right)  =\! - \sqrt{\frac{S^{cl}}{M}}\!.\!\\
\end{split}
\end{equation}

The appropriate boundaries of integration for $\tau_0$ are $-L$ and $L$ since $\tau$ takes values in the interval $(-L,L)$. We then obtain
\begin{equation}
\begin{split}
F = & \mathcal{N} \prod_{n=1}^\infty\! \int_{-\infty}^{\infty}\! \frac{d \chi_n}{\sqrt{2 \pi \hbar /M}}  e^{-\frac{M \lambda_n \chi_n^2}{2 \hbar}} \times \\
& \int_{-L}^L \frac{d \tau_0}{\sqrt{2 \pi \hbar /M}} \left( \sqrt{\frac{S^{cl}}{M}} - \sum_{m=1}^{\infty} \xi_m  r_{m}^\prime (0) \right) \\
= &  \mathcal{N} 2L \sqrt{ \frac{S^{cl} }{2 \pi \hbar}} \frac{1}{\sqrt{\prod _n^\prime \lambda_n}}
 \end{split}
\end{equation}
where $\prod _n^\prime$ indicates the product over the eigenvalues taking out the zero eigenvalue. 

The normalization constant can be conveniently expressed in terms of the sum over harmonic fluctuations around $0$ or $4\pi$. If we define 
\begin{equation}
F_0 = \int [\mathcal{D} \chi] \exp \left(  -\frac{M}{2 \hbar}\int_{-L}^{L} d\tau  \chi D_0  \chi  \right)
\end{equation} 
with 
\begin{equation}
D_0 = - \frac{d^2}{d \tau^2} + \omega^2.
\end{equation}
The normalization constant $\mathcal{N}$ can be written as
\begin{equation}
\mathcal{N} =  F_0  \sqrt{\prod_n \lambda_n^0},
\end{equation}
where $\lambda_n^0$ are the eigenvalues of the differential operator $D_0$. $F_0$, the fluctuation contribution to the imaginary time harmonic oscillator propagator, is readily available in the literature (see e.g. Ref.~\onlinecite{doi:10.1142/9789812562197_0017}). For $L\rightarrow \infty$ its leading contribution is
\begin{equation}
F_0 = \sqrt{\frac{M\omega}{\pi \hbar}} e^{-\omega L}.
\end{equation} 

Our expression for $F$ currently includes a ratio between the products of eigenvalues of the operators $D_0$ and $D$:
\begin{equation}
F =    2L F_0 \sqrt{ \frac{S^{cl} }{2 \pi \hbar}} \sqrt{ \frac{\prod_n \lambda_n^0}{\prod_n^\prime \lambda_n}},
\end{equation}
which can be evaluated using the Gelfand-Yaglom formula. Following Ref.~\onlinecite{doi:10.1142/9789812562197_0017} we have  
\begin{equation}
\frac{\prod_n \lambda_n^0}{\prod _n^\prime \lambda_n} = \frac{2M\omega \eta^2}{S^{cl}},
\end{equation}
where $\eta$ is defined by the asymptotic behavior of the classical solution:
\begin{equation}
\partial_\tau \theta^{cl}  \rightarrow \eta e^{-\omega | \tau|} \quad \text{for } \tau\rightarrow \pm \infty.
\end{equation}

To the leading order the amplitude to propagate from $0$ to $4\pi$ in an imaginary time interval $2L$ is then:
\begin{equation}
(0 , -L | 4\pi , L) \approx 2L   F_0  \eta  \sqrt{ \frac{M \omega}{ \pi \hbar}}e^{ -\frac{S^{cl}}{\hbar}}. 
\end{equation}
However, the leading order contribution is not enough to obtain the level splitting. It is possible to obtain a more precise expression for the amplitude using the dilute instanton gas approximation. 

Under the dilute instanton gas approximation, we sum over paths consisting of combinations of kinks and anti-kinks and quadratic fluctuations around them, i.e. 
\begin{equation}
\theta(\tau) = \sum_{n=0}^{2N} \nu_n \theta^{cl} \left( \tau-\tau_n\right) + \chi(\tau)
\end{equation}
where $\nu_n= \pm1$ ($+$ for kinks and $-$ for anti-kinks) and $\sum_{n} \nu_n =1$. The approximation consist of considering that the centers of the kinks and anti-kinks, i.e. $\tau_n$ are sufficiently spread out to make kink-kink interactions negligible. The obtained result is 
\begin{equation}
\begin{split}
(0 , -L | 4\pi , L) & =\sum_{n} \frac{  F_0 \left(2L \eta  \sqrt{ \frac{M \omega}{ \pi \hbar}} e^{ -\frac{S^{cl}}{\hbar}} \right)\!^{2n+1}}{(2n+1)!} \\
& =  F_0 \sinh \left( 2L \eta  \sqrt{ \frac{M \omega}{ \pi \hbar}} e^{ -\frac{S^{cl}}{\hbar}}  \right).
\end{split}
\label{eqn:kink-antikink}
\end{equation}

The spectral representation of the amplitude in Eq.~\ref{eqn:tun} is 
\begin{equation}
\begin{split}
(0 , -L | 4\pi , L) & =\sum_{n} \psi_n (0) \psi_n (4\pi) e^{-2L E_n/\hbar} .
\end{split}
\end{equation}
Considering two groundstate levels of harmonic oscillators with frequency $\omega$ and mass $M$, one centered around $0$ and other around $4\pi$, which can tunnel to each other with amplitude $\nu$, we have
\begin{equation}
\begin{split}
\psi_1(\theta) = \frac{1}{\sqrt{2}} \left( \psi_0 (\theta) + \psi_{4\pi}(\theta)\right)\quad E_1 = \frac{\hbar \omega}{2} - \nu \\
\psi_2(\theta) = \frac{1}{\sqrt{2}} \left( \psi_0 (\theta) - \psi_{4\pi}(\theta)\right)\quad E_2 = \frac{\hbar \omega}{2} + \nu .
\end{split}
\end{equation}
In the above expression $\psi_0 (\theta)$ and $\psi_{4\pi}(\theta)$ are the groundstate wavefunctions of harmonic oscillators centered around $0$ and $4\pi$, respectively, e.g. 
\begin{equation}
\psi_0 (\theta) = \left(\frac{M\omega}{\pi \hbar}\right)^{1/4} e^{-\frac{M\omega \theta^2}{2\hbar}}.
\end{equation}
The amplitude, Eq.~\ref{eqn:tun}, for such system would then be
\begin{equation}
\begin{split}
(0 , -L | 4\pi , L) & = \sqrt{\frac{M\omega}{\pi \hbar}} e^{-L\omega} \sinh \left(2L \nu/\hbar \right) .
\end{split}
\label{eqn:twolevel}
\end{equation}

Comparing Eq.~\ref{eqn:kink-antikink} and Eq.~\ref{eqn:twolevel} allows us to conclude
\begin{equation}
\nu = \hbar \eta  \sqrt{ \frac{M \omega}{ \pi \hbar}} \exp \left( -\frac{S^{cl}}{\hbar} \right).
\end{equation}

For the kink in Eq.~\ref{eqn:kink} we have:
\begin{equation}
\begin{split}
S^{cl}=&16 M \omega \left(1 + 2R\csch2R \right) \\
\eta =& 8 \omega \cosh R.
\end{split}
\end{equation}
Substituting the values of $M$, $\omega$ and $R$ from Eq.~\ref{eqn:massetc} we obtain
\begin{equation}
\nu_{4\pi}  = \sqrt{\frac{8(\hbar \omega)^5}{\pi E_M E_C^2}}\exp\left(-  \frac{\hbar\omega}{E_C} \times f \left(\frac{E_M}{8E_J} \right) \right)
\end{equation}
with 
\begin{equation}
f \left(x \right) = 2 + \frac{2x}{\sqrt{1+x}}\coth^{-1} \left(\sqrt{1+x} \right).
\end{equation}

\subsection{Emergent translational mode correction for the $4\pi$ phase slip amplitude.}
\label{sec:RApp}

%\textit{Add disclaimer - Follows largely Ref. but since the results were done in the context of statistical mechanics we include them for clarity.}
Here, we follow the procedure outlined in Ref.~\onlinecite{PhysRevB.40.686} to introduce corrections to the previously found expression for $\nu_{4\pi}$. This section then follows the work done in Ref.~\onlinecite{PhysRevB.40.686} closely. We include the calculation here for clarity as the work in Ref.~\onlinecite{PhysRevB.40.686} was done in the context of classical statistical mechanics. We also note that Ref.~\onlinecite{PhysRevB.40.686} claims, incorrectly, that this procedure leads to a non-divergent expression. Here, we find otherwise. 

When $E_M \rightarrow 0$, the expression for $\nu$ found in~\ref{sec:nu4pi} diverges. This occurs because one of the eigenmodes of the operator $D$, which we will call $\lambda_1$ approaches 0 when $E_M \rightarrow 0$. Physically, the two $2\pi$ kinks decouple turning the distance between the two $2\pi$ kinks $2R$ into another translation mode. We must then have
\begin{equation}
y_1(\tau) \rightarrow \partial_R ( \theta^{cl}) \: \text{ when }\: E_M \rightarrow 0 
\end{equation}
This means that we can deal with the effects of the emergent translational mode by writing the path as
\begin{equation}
\begin{split}
\theta(\tau) = & \theta^{cl} (\tau) + \sum_{n=0}^{\infty} \chi_n y_n\left(\tau \right) \\
= & \sigma \left( \tau - \tau_0 , r \right) + \sum_{n=2}^{\infty} \chi_n y_n\left(\tau - \tau_0 \right)
\end{split}
\label{eqn:newpath2}
\end{equation} 
with 
\begin{equation}
\sigma(\tau,r) = 4\arctan[e^{\omega\tau +r}] +  4\arctan[e^{\omega\tau -r}].
\end{equation}
For $R=r$ we recover the classical solution, i.e. $\sigma(\tau,R)=\theta^{cl}(\tau)$ . 
We should note that Eq.~\ref{eqn:newpath2}, and thefore the rest this appendix, relies on $y_1 \approx \partial_R ( \theta^{cl})$. This is a valid assumption when $R>1.25$.\cite{PhysRevB.40.686,PhysRevB.35.3496}

Up to second order in $\chi=\sum_{n=2}^{\infty} \chi_n y_n\left(\tau - \tau_0 \right)=0$ the Lagrangian for the above path is given by:
\begin{equation}
\!\frac{\mathcal{L}\left( \sigma,\chi \right)}{M} \! =\! 
\frac{ \left( \partial_\tau \sigma \!+ \!\partial_\tau \chi \right)^2}{2} +\! V_0 (\sigma) + \!\chi V_1 (\sigma) +\chi^2 V_2(\sigma),\!
\end{equation}
where $V_0 (\sigma) = V(\sigma)$ is the potential of $\sigma$ given by Eq.~\ref{eqn:potential} and 
\begin{equation}
\begin{split}
V_1 (\sigma) = &\frac{\omega^2}{\cosh^2 (R)} \left( \sinh^2 (R) \sin \sigma  + 2\sin \frac{\sigma}{2} \right) \\
V_2 (\sigma) = &\frac{\omega^2}{ \cosh^2 (R)} \left( \frac{1}{2}\sinh^2 (R) \cos \sigma  + \frac{1}{2} \cos \frac{\sigma}{2} \right).\!
 \end{split}
\end{equation}
The action of this path can be written as 
\begin{equation}
S(\sigma,\chi) = S_0 (r)  + S_1(\sigma,\chi)
\end{equation}
with $S_0  (r)$ and $ S_1(\sigma,\chi)$ given by:
\begin{equation}
\begin{split}
S_0 (r)=&  M \int d\tau \left( \frac{1}{2} (\partial_\tau \sigma)^2 + V_0 (\sigma)  \right) \\=& 8 M \omega\left( 1+ \frac{\tanh^2 R}{\tanh^2 r} +\frac{2r}{\sinh 2r} + \right.\\& +\left. \frac{2r\coth r}{\cosh^2 R} - \frac{r\tanh^2 R \coth r}{ \sinh^2 r} \right) \\
S_1(\sigma,\chi) =& M \int d\tau \left( - \partial_\tau \sigma + V_1 (\sigma)  \right) \chi  \\&+M \int d\tau \left( \frac{1}{2} (\partial_\tau \chi)^2 + V_2 (\sigma) \chi^2 \right).  
\end{split}
\end{equation}

Using the Fadeev-Popov method to transform from the coordinates $\chi_0$ and $\chi_1$ to $\tau_0$ and $r$ leads to
\begin{equation}
\begin{split}
(0 , -L |& 4\pi , L) =  \!\mathcal{N} \!\int\!\int\! \frac{d\tau_0 dr }{2 \pi \hbar /M}\! \left.\!\left| \frac{\partial \chi_0 \partial \chi_1}{\partial \tau_0 \partial r }\right| \right|_{ \chi_{0},\chi_{1}=0}  \\
& \prod_{n=2}^{\infty} \int  \frac{d \chi_n}{\sqrt{2 \pi \hbar /M}} e^{-S_0 (r)/\hbar - S_1(\sigma,\chi)/\hbar}
\end{split}
\end{equation}

Following Ref. \onlinecite{PhysRevB.40.686}, we make the approximations:
\begin{equation}
\begin{split}
&\left.\!\left| \frac{\partial \chi_0 \partial \chi_1}{\partial \tau_0 \partial r }\right| \right|_{ \chi_{0},\chi_{1}=0} \!\approx \sqrt{ \!\int \! d\tau (\partial_\tau \sigma )^2  \times \!\int  \!d\tau (\partial_r \sigma )^2 } \\
&\int \prod_{n=2}^{\infty}  \frac{d \chi_n}{\sqrt{2 \pi \hbar /M}} e^{ - S_1(\sigma,\chi)/\hbar}\approx \frac{1}{\sqrt{\prod _n^{\prime \prime} \lambda_n}}
\end{split}
\end{equation}
where the $\lambda_n$s are the eigenmodes of the operator $\mathcal{D}$ from the Eq.~\ref{eqn:D}, and the product $\prod _n^{\prime \prime}$ skips the 0 eigenmode and $\lambda_1$.

Under these approximations, we can write 
\begin{equation}
\begin{split}
(0 , -L | 4\pi , L)  = & F^\prime K 2L 
\end{split}
\end{equation}
with, 
\begin{equation}
F^\prime =\!  \frac{\mathcal{N}}{\sqrt{\prod _n^{\prime \prime} \lambda_n}}\!= F_0 \sqrt{\frac{\prod_n \lambda_n^0}{\prod _n^{\prime \prime} \lambda_n}} = F_0 \eta \sqrt{\frac{2M\omega}{S^{cl}}} \sqrt{\lambda_1} 
\end{equation}
and 
\begin{equation}
\begin{split}
K = \int_{0}^{L\omega} dr & \frac{M  \sqrt{ \!\int \! d\tau (\partial_\tau \sigma )^2  \times \!\int  \!d\tau (\partial_r \sigma )^2 }}{2\pi \hbar} \\
&e^{-S_0(r)/\hbar} \int_{-L+\frac{r}{\omega}}^{L-\frac{r}{\omega}} \frac{d\tau_0}{2L}.
\end{split}
\end{equation}

Using the following result from Ref. \onlinecite{PhysRevB.35.3496}:
\begin{equation}
\begin{split}
\sqrt{ \!\int \! d\tau (\partial_\tau \sigma )^2\!  \times \!\int  \!d\tau (\partial_r \sigma )^2 }  &=\\ &\!16 \sqrt{1-\!4 r^2 \text{csch}^2 2 r}.
\end{split}
\end{equation}
and performing the $\tau_0$ integration gives
\begin{equation}
K\! =\! \int_{0}^{L\omega} \!dr \! \frac{16 M (\omega L\!-\!r) \!\sqrt{1\!-\!4 r^2 \text{csch}^2 2 r}}{2\pi \hbar\omega L}  e^{\frac{-\!S_0(r)}{\hbar}}\! .\!
\label{eqn:K}
\end{equation}

It is possible to calculate $\lambda_1$ by noting that calculating $K$ using a quadratic approximation on $\rho= r-R$ gives
\begin{equation}
K_{0} = \sqrt{\frac{S^{cl}}{2\pi \hbar}} \frac{1}{\sqrt{\lambda_1}} e^{-\frac{S^{cl}}{\hbar}}
\end{equation}
Expanding $ S_0(r)$ up to second order in  $\rho$ leads to
\begin{equation}
\begin{split}
S_0(\rho)= &S^{cl} + 2 M \omega \rho^2 \text{csch}^3R \text{sech}^3R \times \\
&(4 \sinh 2 R-4R+\sinh 4 R-8 R \cosh 2 R)
\end{split}
\end{equation}
We then find 
\begin{equation}
\begin{split}
K_{0} &= \int_{-\infty}^{\infty} d \rho \frac{M 16 \sqrt{1-4 R^2 \text{csch}^2 2R}}{2\pi \hbar} e^{ -\frac{S_0(\rho)}{\hbar}} \\
&= 
\sqrt{\frac{M}{\hbar \omega \pi }} g(R) e^{-\frac{S^{cl}}{\hbar}}
\end{split}
\label{eqn:K0}
\end{equation}
with 
\begin{equation}
g(R)=
\frac{\sinh 2 R \sqrt{2-8 R^2 \text{csch}^2 2 R}}{\sqrt{  \cosh 2 R-R \tanh R-3 R \coth R+2}}
\end{equation}
Note that the factor $(1-\frac{r}{L})$ from Eq.~\ref{eqn:K} goes to 1 in the Eq.~\ref{eqn:K0} as we are taking the $L\rightarrow \infty$ limit. The vanishing eigenvalue $\lambda_1$ is then
\begin{equation}
\begin{split}
&\sqrt{\lambda_1} = \sqrt{\frac{S^{cl}}{2\pi \hbar}} \frac{e^{-\frac{S^{cl}}{\hbar}}}{K_{0}} = \sqrt{\frac{S^{cl} \omega}{2M}} \frac{1}{ g(R)}, 
\end{split}
\end{equation}
which leads to 
\begin{equation}
(0 , -L | 4\pi , L)  =   2L F_0 \frac{\eta\omega}{ g(R)} K  
\end{equation}

Using the dilute instanton gas approximation (see previous section), this result leads to the tunneling amplitude 
\begin{equation}
\nu_{4\pi} = \hbar \frac{\eta\omega}{ g(R)} K =\frac{ 8 \hbar \omega^2 \cosh R}{ g(R)} K  .
\label{eqn:nu}
\end{equation}
Taking $L$ to infinity results in 
\begin{equation}
K = \frac{8 M }{\pi \hbar} \mathcal{I} \left(R,\tfrac{\hbar \omega}{E_C}\right) =\frac{\hbar }{\pi E_C} \mathcal{I} \left(R,\tfrac{\hbar \omega}{E_C}\right) ,
\end{equation}
with $\mathcal{I}(R,\alpha)$ defined by Eq.~\ref{eqn:I}. Our final expression for $\nu$ is 
\begin{equation}
\nu =\frac{ 8 (\hbar \omega)^2 \cosh R}{ g(R) \pi E_C} \mathcal{I} \left(R,\tfrac{\hbar \omega}{E_C}\right) . 
\end{equation}
which corresponds to the expression in the main text Eq.~\ref{eqn:nu4piall} since $F(R) = 2 \cosh R/g(R)$.

\section{Approximate expressions for $\mathcal{I}\left(R, \alpha \right)$ }

\subsection{Validity of the harmonic approximation}
\label{sec:ValSP}

Taking $r = R + y/\sqrt{\alpha S_R^{\prime \prime} (R)}$ we can write the saddle point expansion of $\mathcal{I}\left(R, \alpha \right)$ as
\begin{equation}
\begin{split}
\mathcal{I}\left(R, \alpha \right)   \approx  &\int_{-\infty}^{\infty}  \frac{dy\sqrt{1-4 R^2 \text{csch}^2 2 R} e^{- \alpha S_R (R)}}{\sqrt{\alpha S_R^{\prime \prime} (R)}}\\
&\times e^{- y^2/2}\left( 1 + \sum _{n=1}^\infty \frac{ p_n (y,R)}{(\alpha S_R^{\prime \prime} (R))^{n/2}} \right) \\
= & e^{- \alpha S_R (R)} \sqrt{\frac{2\pi (1-4 R^2 \text{csch}^2 2 R)}{\alpha S_R^{\prime \prime} (R) }} \\ &\times \left( 1 + \sum_{n=1}^\infty \frac{C_n(R)}{(\alpha S_R^{\prime \prime} (R))^n} \right).
\end{split}
\end{equation}
In the above equation the $p_n (y,R)$ are odd/even polynomials in $y$ when $n$ is even/odd, and the $C_n(R)$ are functions of $R$ which can be expressed in terms of derivatives of $S_R(r)$ and $\sqrt{1-4 r^2 \text{csch}^2 2r}$ evaluated at $r=R$.

The expression for $\mathcal{I}\left(R, \alpha \right)$ in Eq.~\ref{eqn:saddleaprox} corresponds to the first term in the above saddle point expansion; therefore, it is a valid approximation if $1/(\alpha S_R^{\prime \prime} (R)) \ll 1$. The function $1/S_R^{\prime \prime} (R) $ diverges for $R \rightarrow 0$ and for $R \rightarrow \infty$ making the approximation for both small and large $R$. However, since Eq.~\ref{eqn:nu4piall} was obtained to address the large $R$ divergence, we only need to find the upper $R$ limit for the validity of Eq.~\ref{eqn:saddleaprox}. Since
\begin{equation}
\frac{1}{\alpha S_R^{\prime \prime} (R)} = \frac{e^{2R}}{16 \alpha } + \mathcal{O} (R),
\end{equation}
Eq.~\ref{eqn:saddleaprox} is valid when $e^{2R} \ll 16 \alpha$. This condition makes the tunneling expression of Eq.~\ref{eqn:nu4piharmonic} valid for $E_M/(8E_J) \gg  E_C/(4 \hbar \omega)$

\subsection{Large \textit{\textbf{R}} limit}
\label{sec:largeR}
To find an approximate expression for $\mathcal{I}\left(R, \alpha \right) $ in the large $R$ limit, we note that $S_R (r)$ grows linearly with $r$ for large $r$. Furthermore, the slope of the large $r$ linear behavior as $R$ increases. This means that, when $R$ is large, the largest contribution to $\mathcal{I}\left(R, \alpha \right) $ will come from the large $r$ linear behavior. We start by writing the following large $r$ expansions: 
\begin{equation}
\begin{split}
 &S_R (r) =  1+ \tanh^2 R+  2r \sech^2 R\\
& + 4e^{-2 r} \left(2r \text{sech}^2 R+\tanh ^2 R\right)+ \mathcal{O} (e^{-4r}) \\
&\sqrt{1-4 r^2 \text{csch}^2 2r} = 1 + \mathcal{O} ( e^{-4r})
\end{split}
\end{equation}
This means we can expand $\mathcal{I}\left(R, \alpha \right)$ as
\begin{equation}
\begin{split}
\mathcal{I}\left(R, \alpha \right) & = \int_{0}^{\infty} dr e^{- \alpha (1+ \tanh^2 R+  2r \sech^2 R)}  \left( 1-\right.\\ \alpha 4e^{-2 r} &\left. \left(2r \text{sech}^2 R+\tanh ^2 R\right) + \mathcal{O} (e^{-4r}) \right)
\\ &= \mathcal{I}_0 \left(R, \alpha \right) + \mathcal{I}_1 \left(R, \alpha \right) + ...
\end{split} 
\end{equation}
where
\begin{equation}
\begin{split}
\mathcal{I}_0\left(R, \alpha \right)  = &\frac{\cosh ^2 R e^{-\alpha  \left(\tanh ^2 R+1\right)}}{2 \alpha } 
\end{split} 
\end{equation}
 corresponds to the approximation to $\mathcal{I}\left(R, \alpha \right) $ cited in Eq.~\ref{eqn:largeRapprox} and $ \mathcal{I}_1 \left(R, \alpha \right)$ is a leading order correction which we calculate to determine the range of validity of Eq.~\ref{eqn:largeRapprox}.

Performing the $r$ interaction gives
\begin{equation}
\begin{split}
\mathcal{I}_1 \!\left(R, \alpha \right) 
= & \frac{-2  \alpha(\alpha  \tanh ^2\! R \text{sech}^2 R\!+\!1)  e^{-\alpha  \tanh ^2\! R-\alpha}}{ \left(\alpha  \text{sech}^2 R\!+\!1\right)^2}\!,\!
\end{split}
\end{equation}
and we obtain:
\begin{equation}
\frac{\mathcal{I}_1 \!\left(R, \alpha \right) }{\mathcal{I}_0 \!\left(R, \alpha \right) } \sim 16 \alpha^2 e^{-2R} \sim 4\alpha^2  \frac{E_M}{8 E_J}.
\end{equation}

The approximation is valid when $16 \alpha^2 e^{-2R} \ll 1$. For $R$ given by Eq.~\ref{eqn:R} and $\alpha = \hbar \omega /E_C$ this is equivalent to $E_M/(8E_J) \ll  0.25 E_C^2/(\hbar \omega)^2$.

For $\mathcal{I}\left(R, \alpha \right) \approx\mathcal{I}_0 \left(R, \alpha \right) $ we obtain 
\begin{equation}
\nu_{4\pi} = \frac{f_2 \left( \frac{E_M}{8 E_J} \right) \left( \hbar \omega \right)^3 }{\pi E_C E_M}\exp \left[ - \frac{\hbar \omega}{E_C} \times f_1 \left( \frac{E_M}{8 E_J} \right) \right]
\end{equation}
with
\begin{equation}
\begin{split}
f_1 (x) =& \frac{2+x}{1+x} \\
f_2 (x) = 
&\left[ \frac{6 (x+1)}{\frac{x}{\sqrt{x+1}} \log \left(\frac{\sqrt{x+1}+1}{\sqrt{x}}\right)+1} \right.\\&+\left.\frac{2}{\frac{x }{\sqrt{x+1}}\log \left(\frac{\sqrt{x+1}+1}{\sqrt{x}}\right)-1} \right]^{1/2}.
\end{split}
\end{equation}

\section{Decoupling of $4\pi$ phase slips}
As it has been previously noted, the expression for $\nu_{4\pi}$ in Eq.~\ref{eqn:4pismallEM} diverges when $E_M\rightarrow 0$. In this appendix, we will show that it is possible to recover the decoupling of the $4\pi$ phase slips into two $2\pi$ phase slips from Eq.~\ref{eqn:nu}. This is achieved by changing the order in which the limits $E_M \rightarrow 0$ and $L\rightarrow 0$ are taken.  

Expanding $F^\prime$ and $K$ around  $x= E_M/(8 E_J)=0$ leads to
\begin{equation}
\begin{split}
F^\prime &=   F_0 \omega^2 \left(4  +  x \left(3-  2\log \left( \frac{x}{4} \right)\right)+ \mathcal{O}(x^2)\right)\\
K&= \frac{4 L M \omega  e^{-\frac{16 M \omega }{\hbar }}}{\pi  \hbar }\\
&-\frac{16 x \left(2L M^2 \omega ^2 (2L \omega -3) e^{-\frac{16 M \omega }{\hbar }}\right)}{3 \left(\pi  \hbar ^2\right)}+\mathcal{O}\left(x^2\right)
\end{split}
\end{equation}
Then, when $E_M \rightarrow 0$, 
\begin{equation}
(0 , -L | 4\pi , L)  \rightarrow  F_0 \frac{8 (2L)^2 M \omega ^3 e^{-\frac{16 M \omega }{\hbar }}}{\pi  \hbar }
\end{equation}
The tunneling amplitude between $0$ and $2\pi$ in a non-topological Josephson junction can be written as
\begin{equation}
 \nu_{2\pi} = 4\omega \sqrt{\frac{\hbar M \omega}{\pi}} e^{-\frac{8 M \omega }{\hbar }}.
\end{equation}
This leads to
\begin{equation}
(0 , -L | 4\pi , L) \rightarrow   F_0 \frac{(2L)^2}{2} \left( \frac{ \nu_{2\pi} }{\hbar} \right)^2
\end{equation}
which is the expected result for propagating between $0$ to $4\pi$ through two uncoupled $2\pi$ phase slips. The $\frac{1}{2}$ factor arises from time ordering the phase slips, i.e.
\begin{equation}
\int_{-L}^{L} d\tau_1 \int_{\tau_1}^{L} d\tau_2 = \frac{(2L)^2}{2}.
\end{equation} 

\ \\ \ \\ \ \\ \ \\ \ \\ \ \\ \ \\ \ \\ \ \\ \ \\ \ \\ \ \\ \ \\ \ \\ \ \\ \ \\

\bibliographystyle{apsrev}
\bibliography{LongPaper2}

\end{document}